\documentclass[pdflatex,sn-basic]{sn-jnl}

\usepackage{graphicx}
\usepackage{amsfonts}
\usepackage{subcaption}
\usepackage{xspace}
\usepackage{booktabs}
\usepackage{multirow}		
\usepackage{mleftright}
\usepackage{hyperref}
\usepackage{siunitx}
\usepackage{thmtools}
\usepackage{amsthm}
\newtheorem{theorem}{Theorem}
\newtheorem{lemma}{Lemma}

\newtheorem{problem}{Problem}

\usepackage{enumitem}
\setlist[enumerate, 1]{labelindent=\parindent, labelwidth=0.5em, leftmargin=!, align=left, topsep=3pt}
\setlist[itemize, 1]{labelindent=\parindent, labelwidth=0.5em, leftmargin=!, align=left, topsep=3pt}

\usepackage[english,linesnumbered,vlined,ruled,nokwfunc]{algorithm2e}
\DontPrintSemicolon
\SetKwComment{note}{$\triangleright$ }{}
\SetFuncSty{textsc}
\SetCommentSty{textit}
\SetDataSty{texttt}
\SetKwInput{Initial}{Initial}
\SetKwInput{Parameter}{Param.}
\SetKwInput{Input}{Input}
\SetKwInput{Data}{Data}
\SetKwInput{Output}{Output}
\ResetInOut{Result} 
\let\oldnl\nl%
\newcommand{\nonl}{\renewcommand{\nl}{\let\nl\oldnl}}%
\SetKw{KwAnd}{and} %
\SetKw{KwOr}{or}
\SetKw{KwXor}{xor}
\SetKw{KwNot}{not}
\SetKw{Parallel}{parallel}
\SetKw{Return}{return}
\SetKw{Break}{break}

\usepackage{dsfont}
\newcommand{\id}[1]{\mathds{1}_{\{#1\}}}

\usepackage{tikz}
\usetikzlibrary{arrows,decorations.pathmorphing}
\usetikzlibrary{arrows.meta}
\usepackage{tikz-network}

\usepackage{cleveref}

\crefname{problem}{problem}{problems}
\Crefname{problem}{Problem}{Problems}

\usepackage[table,dvipsnames,svgnames]{xcolor} 

\newcommand{\alg}{\texttt{Cosmic}\xspace}
\newcommand{\algp}{\texttt{Cosmic+}\xspace}
\newcommand{\algvg}{\texttt{Cosmic(G)}\xspace}
\newcommand{\algpvg}{\texttt{Cosmic+(G)}\xspace}
\newcommand{\algvp}{\texttt{Cosmic(P)}\xspace}
\newcommand{\algpvp}{\texttt{Cosmic+(P)}\xspace}

\newcommand{\revision}[1]{\textcolor{black}{#1}}

\newcommand{\ecr}{\kappa_e}

\newcommand{\wecr}{\kappa_w}

\begin{document}

\title{Consistent Tie-Strength Labeling for Multilayer Strong Triadic Closure}

\author*[1]{\fnm{Lutz} \sur{Oettershagen}}\email{lutz.oettershagen@liverpool.ac.uk}

\author[2]{\fnm{Athanasios L.} \sur{Konstantinidis}}\email{a.konstantinidis@uoi.gr}

\author[3]{\fnm{Fariba} \sur{Ranjbar}}\email{fariba.ranjbar@luiss.it}

\author[3]{\fnm{Giuseppe F.} \sur{Italiano}}\email{gitaliano@luiss.it}

\affil[1]{\orgname{University of Liverpool}, \orgaddress{\city{Liverpool}, \country{UK}}}

\affil[2]{\orgname{University of Ioannina}, \orgaddress{\city{Ioannina},  \country{Greece}}}

\affil[3]{\orgname{Luiss University}, \orgaddress{\city{Rome}, \country{Italy}}}

\abstract{
Inferring tie strengths (\emph{strong} vs.~\emph{weak}) is a core task in network analysis, often guided by the Strong Triadic Closure (STC) principle. In multilayer networks, such as social platforms or biological systems, applying STC independently to each layer can lead to inconsistent tie labels, undermining interpretations that rely on coherent relationship semantics across layers.
We propose new formulations, multilayer STC and its extension STC+, which are axiomatically grounded and enforce cross-layer consistency.
These problems are NP-hard; we present efficient 2- and 6-approximation algorithms alongside exact solutions. Experiments on real-world networks demonstrate that our methods produce consistent tie strength labelings with a transparent structural justification, significantly improving over the baselines.
}

\keywords{strong triadic closure, multilayer networks, tie strength}

\maketitle

\section{Introduction}
\revision{
Inferring the strength of relationships between entities is a fundamental task in network analysis, with applications in social network mining and related interaction networks~\citep{adriaens2020relaxing,GilbertK09,kahanda2009using,lu2010link,oettershagen2024inferring,RozenshteinTG2017,stolz2021predicting}.
In many real-world networks, edges are not equally meaningful: some represent \emph{strong} (close, persistent) interactions, while others capture incidental or \emph{weak} relationships~\citep{granovetter1973strength}. 
Distinguishing between \emph{strong} and \emph{weak} ties is therefore essential for understanding network structure and for supporting downstream tasks such as recommendation, community detection, or robustness analysis.}

\begin{figure}
	\begin{subfigure}{0.3\linewidth}\centering
		\includegraphics[width=.9\linewidth]{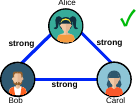}  
		\caption{All three are strongly connected.}
		\label{fig:stc:a}
	\end{subfigure}\hfill%
	\begin{subfigure}{0.3\linewidth}\centering
		\includegraphics[width=.9\linewidth]{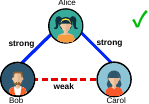}
		\caption{We can have one, two, or three weak edges.}
		\label{fig:stc:b}
	\end{subfigure}\hfill%
	\begin{subfigure}{0.3\linewidth}\centering
		\includegraphics[width=.9\linewidth]{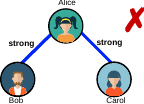}
		\caption{An open wedge with two strong edges is not allowed.}
		\label{fig:stc:c}
	\end{subfigure}
	\caption{\revision{Illustration of the \emph{Strong Triadic Closure (STC)} principle.
			Alice is strongly connected to Bob and Carol. In (a), all three edges are strong and the STC is satisfied. In (b), Bob and Carol are connected by a weak edge, which still satisfies the STC. In (c), Bob and Carol are not connected, resulting in an open wedge with two strong edges and thus a violation of the STC.}}
	\label{fig:stc}
\end{figure}

\revision{A prominent topology-based principle for tie strength inference is the \emph{Strong Triadic Closure} (STC) property, introduced by \cite{sintos2014using}. 
Informally, STC states that if a node has strong ties to two others, then those two nodes should be at least weakly connected. 
That is, if person $A$ is strongly connected to $B$, and $B$ is strongly connected to $C$, then $A$ and $C$ should know each other.
This captures the intuition that strong relationships tend to close triangles. \Cref{fig:stc} shows an example.
A key advantage of the STC is that we are able to infer tie strength using only network topology, without requiring node or edge attributes, which are often unavailable due to data collection limitations or privacy concerns.}

\revision{Therefore, \cite{sintos2014using} introduced optimization formulations of the STC problem that seek to label edges as \emph{strong} or \emph{weak}, using only the network structure, so as to satisfy the strong triadic closure constraints. In addition to the basic STC formulation, they proposed an additional variant, denoted STC+, in which new edges may be inserted to resolve violations of the STC, allowing a more flexible trade-off between edge insertions and weak tie assignments.}

\revision{In this work, we introduce new definitions and algorithms to compute the STC and STC+ variants in multilayer networks. In the following we motivate this new direction and discuss the arising challenges as well as our contributions in solving them.}

\begin{figure}
	\centering
	\includegraphics[width=0.4\linewidth]{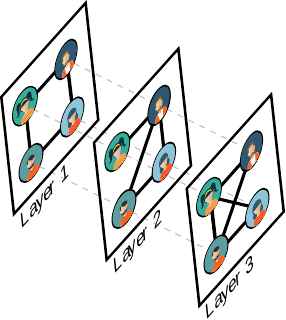}
	\caption{\revision{A \emph{multilayer network} with three layers over a common node set.
			Each layer represents a different interaction type and may have a different edge set. We assume a \emph{single latent tie strength for each node pair}, whose semantics are shared across all layers, independently of edge existence. This motivates consistent tie-strength inference across layers.}}
	\label{fig:multilayer_example}
\end{figure}

\paragraph{Multilayer networks.}
\revision{Many complex systems are more naturally modeled not as a single graph, but as a collection of graphs over the same set of nodes, each representing a different type of interaction.
Such systems are commonly referred to as \emph{multilayer networks}\footnote{\revision{Here, the term ``layer'' refers to an interaction layer and should not be confused with layers in neural-network models.}}~\citep{de2023more,de2013mathematical,kivela2014multilayer}.
Typical examples include social networks combining online and offline interactions, biological networks integrating multiple types of molecular interactions, or collaboration networks where each layer corresponds to a different venue or context.
Unlike temporal snapshots, where layers represent the same relation observed at different times, multilayer networks capture {distinct but coexisting types of relationships}. \Cref{fig:multilayer_example} shows an example of a 3-layer network.}

\revision{We strictly focus on multilayer networks with \emph{semantically aligned layers}, where different layers capture interactions in distinct contexts or modalities but are intended to support a \emph{single} notion of tie strength.
We model this by assuming that each dyad that is observed in at least one layer is associated with one latent binary tie strength (\emph{strong} vs.\ \emph{weak}), whose semantics are shared across layers.
Individual layers provide complementary, potentially noisy and incomplete evidence about this latent tie strength: in particular, the absence of an edge in a layer is not interpreted as ``weak'', but merely as missing or unobserved interaction evidence in that layer.
Equivalently, we view the layers as partial observations that provide \emph{hints} about an underlying binary tie strength, rather than as carrying layer-specific strength semantics.}

\begin{figure}\centering
	\begin{subfigure}{0.45\linewidth}\centering
		\includegraphics[width=.8\linewidth]{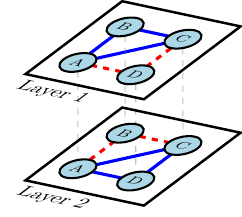}
		\caption{Inconsistent labeling}
		\label{fig:introduction:a}
	\end{subfigure}\qquad%
	\begin{subfigure}{0.45\linewidth}\centering
		\includegraphics[width=.8\linewidth]{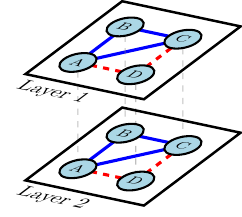}
		\caption{Consistent labeling}
		\label{fig:introduction:b}
	\end{subfigure}
	\caption{Illustration of inconsistent versus consistent edge labelings in a two-layer network. Strong edges (\textcolor{blue}{solid blue}) and weak edges (\textcolor{red}{dashed red}) represent inferred tie strengths. While both (a) and (b) satisfy the STC property in each layer, only (b) maintains consistent edge semantics across layers.}
	\label{fig:introduction}
\end{figure}

\paragraph{The challenge of consistency.}
\revision{A straightforward way of applying STC to multilayer networks is to solve the STC problem independently in each layer.
This produces a \emph{strong/weak} labeling for every layer, but can easily lead to \emph{inconsistent tie strength semantics}: the same edge may be inferred as strong in one layer and weak in another.
Although each layer-wise labeling may satisfy STC in isolation, such inconsistencies undermine interpretability and complicate cross-layer analysis.
\Cref{fig:introduction} illustrates this phenomenon in a two-layer network.
While both labelings in \Cref{fig:introduction:a,fig:introduction:b} satisfy STC within each layer, only the labeling in \Cref{fig:introduction:b} preserves consistent edge semantics across layers.
Moreover, enforcing the STC on an aggregated graph (e.g., merging all layers into a single weighted network in which the weights correspond to the multiplicities of the edges in the multilayer graph) is generally insufficient for guaranteeing STC validity in each layer, since a wedge that is open in a particular layer may appear closed in the aggregated graph due to an edge that exists only in a different layer.\looseness=-1}

\revision{Therefore, our goal is to infer a single, coherent assignment of tie strengths that simultaneously satisfies the STC constraints induced by all layers and preserves a consistent semantic interpretation across layers.}

\paragraph{When is consistency appropriate?}
\revision{We emphasize that cross-layer consistency is a \emph{modeling assumption}, not a universal requirement.
It is well suited for multilayer networks in which layers provide complementary views of the same underlying dyadic relationships, and where the meaning of tie strength (e.g., close vs.~distant) is expected to remain stable across interaction types.
Typical examples include social networks combining digital and physical interactions, as well as biological networks in which multiple experimental layers capture complementary aspects of the same underlying molecular interactions.
In such settings, inconsistent tie strength labels across layers can distort the interpretation of relationships and hinder meaningful cross-layer analysis.
Accordingly, we model tie strength as a binary latent variable (\emph{strong} or \emph{weak}) whose semantics are preserved across layers.}

\revision{In contrast, in highly heterogeneous multilayer systems, such as networks combining family ties and professional collaborations, it may be entirely reasonable for a relationship to be strong in one layer and weak in another.
Our framework is designed for the former class of networks, where consistent semantics are meaningful.}

\paragraph{Our approach.}
\revision{To avoid ad hoc design choices, we adopt an axiomatic approach to formulating STC in multilayer networks.
We identify five natural axioms that any reasonable multilayer STC objective should satisfy: 
(1) penalization of weak edges, 
(2) layer permutation invariance, 
(3) edge-wise decomposability, 
(4) non-decreasing disagreement penalties with more strong labels, and 
(5) invariance to duplicating weak-only layers.
These axioms characterize a family of admissible objective functions parameterized by a non-decreasing weighting function and lead to principled optimization problems for multilayer STC and its extension STC+, which allows edge insertions.}

\revision{Both problems are NP-hard.
We develop efficient approximation algorithms with constant-factor guarantees and provide exact integer and linear programming formulations.
Our methods infer a single, coherent labeling of tie strength that simultaneously respects STC constraints in all layers.}

\paragraph{Contributions.}
\revision{Our main contributions are:
\begin{itemize}
	\item We introduce the first \emph{axiomatically grounded} formulations of the STC and STC+ problems for multilayer networks, ensuring consistent tie strength semantics across all layers.
	\item We design efficient approximation algorithms with provable guarantees: a 2-approximation for multilayer STC and a 6-approximation for multilayer STC+, along with exact integer and linear programming formulations.
	\item We conduct extensive experiments on real-world multilayer networks from diverse domains, demonstrating that layer-wise baselines produce inconsistent and often misleading labelings, while our approach yields fully consistent and interpretable results.
\end{itemize}}

\section{Related Work}

There are several recent surveys on multilayer networks, e.g.,~\cite{boccaletti2014structure,crainic2022taxonomy,de2013mathematical,DickisonMagnaniRossi2016,hammoud2020multilayer}.
\cite{pappalardo2013measuring} measure tie strength in multilayer networks, where a greater number of connections across different layers corresponds to a higher (continuous) tie strength between two users. However, their work does not consider the STC property.
Other studies have predicted tie strength based on node and edge features in conventional networks, e.g.,~\cite{GilbertK09,jones2013inferring,PhamSL16,XiangNR10}, but these approaches also do not classify edges with respect to the STC.
In contrast, our work is grounded in the STC framework, and we infer binary tie strengths in each layer purely from the network topology. \cite{sintos2014using} introduced the STC optimization problems, characterizing edges as strong or weak based solely on structural information. They proved that maximizing the number of strong edges is NP-hard and provided two approximation algorithms. An extensive analysis of STC can be found in \cite{EasleyK2010}.
Subsequent work has explored restricted network classes to further analyze the complexity of STC maximization~\citep{GruttemeierK20,KonstantinidisNP18,KonstantinidisP20}. \cite{RozenshteinTG2017} incorporated community connectivity constraints, while \cite{adriaens2020relaxing} proposed integer linear programming formulations and relaxations.
More recently, \cite{MatakosG22} studied the problem of reinforcing connectivity by adding new edges guided by strong ties.
The STC framework has also been extended to dynamic and temporal settings. \cite{oettershagen2022inferring} considered temporal networks with evolving topology using a sliding time window, and later extended this approach to the STC+ variant~\citep{oettershagen2024inferring}. \cite{wickrama2024dense} combined the STC with the densest subgraph problem, while \cite{veldt2022correlation} examined connections between STC+ and cluster editing.

To the best of our knowledge, our work is the first to address the STC in multilayer networks. We ground our formulation in a set of explicit axioms that capture essential principles any consistent tie strength inference method should satisfy.

\section{Preliminaries}

\Cref{table:notation} in the appendix gives an overview of the notation and symbols.
We use $[k]$ with $k \in \mathbb{N}$ to denote the set $\{1,\ldots,k\}$.

\smallskip
\noindent
\textbf{Multilayer Graphs, and (Hyper-)graphs:} 
We define a \emph{multilayer graph} $G=(V, E_1, \ldots, E_k)$ consisting of a finite set of nodes $V$ and $k$ finite sets $E_i$ with $i\in[k]$ of undirected edges $e=\{u,v\}$ with $u$ and $v$ in $V$, and $u\neq v$. 
We define $n=|V|$, $m_i=|E_i|$, and $m=\sum_{i=1}^{k}m_i$.
We call a $1$-layer graph just a \emph{graph}; e.g., each individual layer $G_i = (V, E_i)$, with $i \in [k]$, of a multilayer graph $G = (V, E_1, \ldots, E_k)$ is itself a graph.
Given a graph $G=(V,E)$, we define a \emph{wedge} as a triplet of nodes $u,v,w \in V$ such that $\{\{u,v\},\{v,w\}\}\subseteq E$ and $\{u,w\}\notin E$. We denote such a wedge by $(v,\{u,w\})$, and with $\mathcal{W}(E)$, the set of wedges in the set of edges $E$. Moreover, $\mathcal{W}(E,\{u,w\})=\{(v,\{u,w\})\mid (v,\{u,w\})\in \mathcal{W}(E)\} $ denotes the set of wedges w.r.t.~$\{u,w\}$. 
\revision{Note that $W(E,\{u,w\})=\emptyset$ whenever $\{u,w\}\in E$, since wedges are defined
	only with respect to missing edges.}
A \emph{hypergraph} $H=(V,E)$ consists of a finite set of nodes $V$ and a finite set of \emph{hyperedges} $E\subseteq 2^V\setminus\emptyset$, i.e., each hyperedge connects a non-empty subset of~$V$. 
We consider the special case of \emph{3-uniform} hypergraphs in which each hyperedge consists of an unordered triple. 
We define the union $H= H_1\cup H_2$ of two (hyper-)graphs $H_1=(V_1,E_1)$ and $H_2=(V_2,E_2)$ as $H=(V_1\cup V_2, E_1\cup E_2)$.

Given a graph $G = (V,E)$, we assign each edge $e \in E$ one of the labels \emph{weak} or \emph{strong}. Such a labeling is called a \emph{strong-weak labeling} and is represented by a subset $S \subseteq E$, where each edge $e \in S$ is labeled \emph{strong}, and each $e \in E \setminus S$ is labeled \emph{weak}.

The \emph{strong triadic closure (STC)} property requires that for any two strong edges $\{u,v\} \in S$ and $\{v,w\} \in S$, the edge $\{u,w\}$ must also exist in $E$ (either weak or strong). We say that a labeling $S \subseteq E$ \emph{fulfills} the STC if this condition holds.
Equivalently, the STC is fulfilled if and only if no wedge in $\mathcal{W}(E)$ contains more than one strong edge; that is, there is no triple $(v,\{u,w\})$ such that both $\{u,v\}$ and $\{v,w\}$ are strong and $\{u,w\} \notin E$; cf.~\cite{sintos2014using}.
Two natural optimization variants have been defined and studied:
\begin{itemize}
	\item \textsc{MaxSTC}: maximize $|S|$ subject to $S \subseteq E$ fulfilling STC.
	\item \textsc{MinSTC}: minimize $|E \setminus S|$ subject to $S \subseteq E$ fulfilling STC.
\end{itemize}
\revision{These two formulations are complementary: they optimize over the same feasible set of STC-valid labelings, but with opposing objectives, i.e., maximizing the number of strong edges versus minimizing the number of weak edges. As a result, an optimal solution to one formulation can be obtained from an optimal solution to the other. However, the two problems differ substantially from an algorithmic perspective: approximation guarantees do not carry over, since they are defined with respect to different objective values. \textsc{MinSTC} admits a $2$-approximation, while \textsc{MaxSTC} is hard to approximate within $n^{1-\epsilon}$ for any constant $\epsilon > 0$ unless P = NP~\citep{sintos2014using}.}

\smallskip
\noindent
\textbf{Approximation in Single-Layer Graphs:}  
To approximate \textsc{MinSTC} in a single-layer graph $G=(V,E)$, we first construct the wedge graph $W(G)=(V_W,E_W)$.
The wedge graph $W(G)$ contains for each edge $\{i,j\}\in E$ one vertex $n_{ij}\in V_W$. Two vertices $n_{uv}, n_{uw}$ in $W(G)$ are adjacent if and only if there exists a wedge $(u,\{v,w\})\in \mathcal{W}(E)$. 
\cite{sintos2014using} showed that solving the \emph{vertex cover (VC)} problem on $W(G)$ leads then to a solution for the STC of $G$, where VC is defined as follows: Given a graph $G=(V, E)$, a minimum vertex cover is a subset $C\subseteq V$ such that each edge $e\in E$ is incident to at least one vertex $v\in C$.
Following the definition of $W(G)$, choosing the set of \emph{weak} edges to be the edges $\{u,v\}\in E$ such that $n_{uv}\in C$, each wedge in $\mathcal{W}(E)$ has at least one \emph{weak} edge, i.e., the labeling fulfills the STC.

Recently, the approximation was generalized for edge weighted graphs~\citep{oettershagen2022inferring} by mapping the edge weights of the input graph to node weights in the wedge graph, and solving the \emph{Minimum Weighted Vertex Cover (MWVC)} problem.  The running time of this approximations is in $\mathcal{O}(nm)$ due to the size of the wedge graph.
We adopt and extend this wedge-graph-based approach to approximate the multilayer STC problem, incorporating additional structural constraints across layers.

\smallskip
\noindent
\textbf{Strong Triadic Closure with Edge Additions (STC+):}
In the STC+ variant, apart from labeling the edges of the graph as \emph{strong} or \emph{weak}, new \emph{weak} edges between non-adjacent nodes can be added. The \textsc{MinSTC+} problem is defined as follows~\citep{sintos2014using}:
Given a graph $G = (V,E)$, the goal is to find a set $F \subseteq \binom{V}{2}\setminus E$ and a set $E' \subseteq E$ such that the strong edge set $E \setminus E'$ fulfills the strong triadic closure with respect to the augmented graph $(V, E \cup F)$, while minimizing $|E' \cup F|$.

STC+ is useful, for example, when a single missing edge can force many others to be labeled weak under STC~\citep{sintos2014using}. Adding such edges can significantly reduce the number of weak labels required.

\revision{In previous work, we}
extended the approximation of the STC to the STC+ problem~\citep{oettershagen2024inferring}. Given a graph $G$, we construct a 3-uniform wedge hypergraph in which nodes represent edges of the input graph or edges that can be newly inserted. Hyperedges correspond to the wedges in $G$.
We obtain an approximation solving a MWVC analogously to the approximation of the weighted \textsc{MinSTC} problem; the running time is in $\mathcal{O}(nm)$.
Our algorithm for the multilayer STC+ builds on this idea and generalizes it to multilayer networks.

\section{Strong Triadic Closure in Multilayer Graphs}\label{sec:mlstc}
Our goal is to find an STC labeling for each layer that is consistent across the multilayer network. Formally defining a cost function that properly balances the penalization of \emph{weak} edges with disagreements across layers is a non-trivial task. To ensure our formulation is robust, interpretable, and free from arbitrary design choices, we adopt an axiomatic approach. We define the following five axioms that any reasonable multilayer STC objective should fulfill:

\begin{enumerate}
	\item \emph{Weakness Minimization:} The cost function must include a cost for every edge instance labeled \emph{weak}.
	\item \revision{\emph{Layer Permutation Invariance:} The final cost of a labeling must be independent of the order in which the layers are presented.}
	\item \emph{Edge-wise Decomposability:} The total cost of a labeling for the entire multilayer network can be expressed as the sum of costs calculated for each individual edge.
	\item \emph{Strong Persistence Penalty:} If and only if an edge has conflicting labels (is \emph{strong} in at least one layer and \emph{weak} in at least one), it must incur a disagreement penalty. This penalty must not decrease as the number of layers labeling it \emph{strong} increases.
	\item \emph{Weak Duplication Invariance:} Duplicating a layer where an edge is already known to be \emph{weak} (or is absent) must not alter the optimal resolution for that edge. 
\end{enumerate}

Axiom (1) 
anchors the objective in the classical MinSTC tradition: strong ties are the scarce resource and every weak instance carries budgetary cost.
Axiom (2) 
rules out artifacts from data representation: permuting, batching, or reordering layers must not change the value of any labeling.
Axiom (3) 
requires separability across edges. It lets us write the global objective as a sum of per-edge terms plus feasibility via STC constraints. 
%
Axiom (4) 
captures the semantics of \emph{strong}: if an edge is called \emph{strong} somewhere and \emph{weak} elsewhere, that disagreement must be penalized, and the penalty should not decrease as the number of \emph{strong} occurrences grows. Intuitively, each \emph{strong} claim creates closure obligations in its layer; disagreeing against more such claims should never be cheaper.
Axiom (5) 
prevents gaming by data replication. Duplicating a layer in which an edge is \emph{weak} (or absent) must not alter which labelings are optimal for that edge. In particular, disagreement penalties must not grow with the number of \emph{weak} occurrences, otherwise duplicating layers that contain only weak edges would unduly bias the optimizer's outcome.

We show that these axioms naturally lead to a specific cost objective.
In the following, we use $E=\bigcup_{i\in [k]} E_i$ and for each $e\in E$ we define $s(e)$ to be the number of layers in which $e$ is labeled \emph{strong} and $w(e)$ the number of layers in which $e$ is labeled \emph{weak}. 

\begin{theorem}\label{theorem:costfunction}
\revision{Any cost function that assigns constant costs (without loss of generality, unit costs) satisfies Axioms~(1)--(5) if and only if it has the form}
	$$\text{Cost}(S_1, \ldots, S_k) = \sum_{e \in E} f(s(e), w(e))$$
	where $f : \mathbb{N}_0 \times \mathbb{N}_0 \to \mathbb{R}_{\geq 0}$ satisfies:
	\begin{enumerate}
		\item[(i)] $f(s(e), w(e)) \geq w(e)$ for all $s(e), w(e)$,
		\item[(ii)] $f(s(e), 0) = 0$ for all $s(e)$ (no cost if edge is consistently strong),
		\item[(iii)] $f(s(e), w(e)) = w(e) + g(s(e))$ for some non-decreasing function $g : \mathbb{N}_0 \to \mathbb{R}_{\geq 0}$ with $g(0) = 0$ when $w(e) > 0$.
	\end{enumerate}
\end{theorem}

Guided by Axioms (1)-(5) and \Cref{theorem:costfunction}, we now formalize the multilayer STC objective. \Cref{theorem:costfunction} characterizes the class of valid cost functions under the axioms but leaves the 
disagreement penalty function $g$ undetermined. 
We first define \emph{disagreements} between per-layer labelings.
We say that labelings $S_i$ and $S_j$, with $i\neq j\in[k]$, \emph{disagree} on edge $e$ if exactly one of them labels $e$ strong, i.e., $e\in S_i\cup S_j$ and $e\not \in S_i\cap S_j$.
We then define the \emph{disagreement cost} of the labelings $S_1,\ldots, S_k$ as 
\[
d_k(S_1,\ldots,S_k)=\sum_{\substack{e\in E\\w(e)>0}}s(e).
\]
A labeling is \emph{consistent} if and only if $d_k(S_1, \dots, S_k) = 0$.
We can now define the STC problem in multilayer graphs as follows.
\begin{problem}[\textsc{MinMultiLayerSTC}]
	Given a multilayer graph $G=(V, E_1, \ldots, E_k)$, find labelings $S_i\subseteq E_i$ for $i\in[k]$ such that 
	\begin{enumerate}
		\item for each $i\in [k]$, $S_i$ satisfies the STC for $E_i$, and
		\item $
		\text{Cost}(S_1, \ldots, S_k) = \sum_{i\in[k]} |E_i \setminus S_i| + d_k(S_1,\ldots,S_k) \text{~~is minimized.}
		$
	\end{enumerate}
\end{problem}

This objective directly instantiates \Cref{theorem:costfunction} with the linear disagreement penalty $g(s(e)) = s(e)$ when $w(e) > 0$, which is the minimal non-decreasing function satisfying Axiom~(4). Each \emph{strong} occurrence of an edge that is not consistently labeled across layers contributes one unit to the disagreement penalty.

The \textsc{MinMultiLayerSTC} problem is a generalization of the single-layer \textsc{MinSTC}. 
Moreover, by \emph{layer anonymity} and \emph{edge-wise decomposability} (Axioms~(2)-(3)), any disagreement surcharge for an edge $e$ can depend only on its occurrence counts $(s(e),w(e))$, not on the order of layers.
\emph{Weak Duplication Invariance} (Axiom~(5)) rules out any dependence on $w(e)$ beyond the event $w(e)>0$ to prevent that duplicating a weak-only layer would spuriously change the objective.
\emph{Strong Persistence Penalty} (Axiom~(4)) then requires the surcharge to be nondecreasing in $s(e)$.
Consequently, we penalize \emph{each strong occurrence} of $e$ whenever $e$ is not uniformly labeled across layers.
Operationally, each strong claim creates STC closure obligations in its layer; disagreeing with more such claims should never be cheaper.

Note that solving the single-layer STC optimally in each layer does not lead to an optimal solution for the multilayer variant. For example, in \Cref{fig:introduction:a} each layer is optimally solved separately; however, the optimal solution for the multilayer STC is shown in \Cref{fig:introduction:b}.
\revision{Furthermore, for \textsc{MinMultiLayerSTC}, the STC constraints are evaluated \emph{within each layer}: if $\{u,v\}$ and $\{v,w\}$ are labeled strong in layer $i$, then the closing edge $\{u,w\}$ must be present in $E_i$ (weak or strong) to avoid an open wedge in that layer. Edges that exist only in other layers do not close wedges in layer $i$.
Hence, enforcing STC on an aggregated union graph is generally insufficient for guaranteeing STC validity in each layer.}

Finally, a maximization variant generalizing \textsc{MaxSTC} can be defined, but it inherits the hardness of approximation. Thus, here we focus on the minimization variant, which admits constant-factor approximations.

In the remainder of this section, we give a 2-approximation for \textsc{MinMultiLayerSTC}.
The following lemma provides the basis for our approximation algorithm and the ILP formulation. It establishes that we can focus on finding consistent solutions without loss of optimality.
Intuitively, it states that if there is an optimal solution with disagreements, then there exists also an optimal solution without disagreements. 

\begin{lemma}\label{lemma:eqgeneral} 
	Let $S_1^*, \ldots, S_k^*$ be labelings of strong edges for a multilayer graph $G=(V, E_1, \ldots, E_k)$, where $S_i^* \subseteq E_i$ for each $i \in [k]$. Let $C^*$ be their objective value for the \textsc{MinMultiLayerSTC} problem, i.e., 
	$C^* = \sum_{i\in[k]}|E_i \setminus S_i^*| + d_k(S_1^*,\ldots,S_k^*)$.
	There exists a set of labelings of strong edges $H^*=\{H_1^*, \ldots, H_k^*\}$ (where $H_i^* \subseteq E_i$) s.t.:
	\begin{enumerate}
		\item The labelings in $H^*$ are consistent, i.e., $d_k(H_1^*, \ldots, H_k^*) = 0$.
		\item The objective value of $H^*$ for \textsc{MinMultiLayerSTC} is equal to $C^*$, i.e., $\sum_{i\in[k]}|E_i \setminus H_i^*| + d_k(H_1^*, \ldots, H_k^*) = C^*$.
	\end{enumerate}
\end{lemma}

\Cref{lemma:eqgeneral} is crucial because it allows us to simplify the search space for \textsc{MinMultiLayerSTC}. Instead of needing to find a labeling that might have disagreements and then account for the $d_k$ penalty, we can seek an optimal solution that is inherently consistent (i.e., $d_k(S_1,\ldots,S_k)=0$). In such a consistent labeling, an edge $e$ is either \emph{strong} in all layers it appears in, or \emph{weak} in all layers it appears in. 

Our approximation algorithm for \textsc{MinMultiLayerSTC}, presented as \Cref{alg:klayerstc}, is designed based on this principle. 
It first constructs individual wedge graphs $W_i$ for each layer $G_i=(V,E_i)$. These are then combined into a single, node-weighted wedge graph $W=(V_W, E_W)$. The nodes $V_W$ in $W$ correspond to the unique edges $e \in E$ present in the input multilayer graph. The weight $\omega(n_e)$ of a node $n_e \in V_W$ (representing edge $e$) is set to the multiplicity of $e$. 
An edge $(n_{e_1}, n_{e_2})$ exists in $E_W$ if $e_1$ and $e_2$ form two sides of a wedge in some layer $G_j$. 
The algorithm then finds an approximate Minimum Weighted Vertex Cover (MWVC) in $W$. Edges corresponding to nodes in this MWVC are labeled \emph{weak} across all their layers, yielding a consistent STC labeling.\looseness=-1

\begin{theorem}\label{theorem:approxstc}
	Given an $\alpha$-approximation for the MWVC with running time $T_{\text{VC}}$,
	\Cref{alg:klayerstc} computes an $\alpha$-approximation for the \textsc{MinMultiLayerSTC}
	in 
	$\mathcal{O}(k\cdot nm+T_{\text{VC}})$ running time.
\end{theorem}

\begin{algorithm}
	\caption{A 2-approximation algorithm for \textsc{MinMultiLayerSTC}}\label{alg:klayerstc}
	\Input{Multilayer graph $G=(V,E_1,\ldots,E_k)$}
	\Output{Consistent labelings $S_1,\ldots,S_k$}
	$V_W \gets \emptyset$; $E_W \gets \emptyset$ \tcp*{Initialize combined wedge graph $W$}
	\For{$i\in \{1,\ldots,k\}$}{
		Compute wedge graph $W_{i}=(V_i^\text{wg}, E_i^\text{wg})$ of $G_i=(V, E_i)$\;
		$V_W \gets V_W \cup V_i^\text{wg}$ \tcp*{Nodes in $V_i^\text{wg}$ are $n_e$ for $e \in E_i$}
		$E_W \gets E_W \cup E_i^\text{wg}$\;
	}
	\tcp*[l]{For each unique edge $e \in E = \bigcup E_j$ represented by $n_e$:}
	\For{$n_e\in V_W$}{
		$\omega(n_e)\gets|\{j \mid e \in E_j\}|$\tcp*{Weight is multiplicity of edge $e$}
	}
	Approx.~MWVC $C \subseteq V_W$ on $W$ with node weights $\omega$\label{alg:klayerstc:mwvc_step}\;
	$E_\text{weak} \gets \{ e \mid n_e\in C\}$\tcp*{Set of unique edges labeled weak}
	\For{$i\in[k]$}{
		$S_i\gets E_i\setminus E_\text{weak}$ \tcp*{Strong labeled edges in layer $i$}
	}
	\Return{strong edge sets $S_i$ for each layer $i\in [k]$}
\end{algorithm}

The MWVC solver used in Line~\ref{alg:klayerstc:mwvc_step} of \Cref{alg:klayerstc} can be chosen flexibly. For example, employing the \emph{pricing algorithm} yields a $2$-approximation for MWVC with running time linear in the number of edges of the wedge graph~\citep{bar1981linear}. 
By \Cref{theorem:approxstc}, this implies an overall $2$-approximation for \textsc{MinMultiLayerSTC} with running time $\mathcal{O}(k \cdot n m)$, where $n = |V|$, $m = \sum_{i=1}^k |E_i|$ denotes the total number of edges across all layers, and $k$ is the number of layers.

\revision{Regarding practical applicability, the computational bottleneck of our approach lies in enumerating wedges across layers, making scalability primarily dependent on local density and clustering. In our experimental evaluation (\Cref{sec:experiments}), we demonstrate scalability to large empirical multilayer networks with hundreds of thousands of nodes and over a million edges, while extremely dense or web-scale networks would require additional sparsification or parallelization.}

Moreover, the space complexity of \Cref{alg:klayerstc} is dominated by storing the wedge graph $W$, which requires $\mathcal{O}(knm)$ space in the worst case.
Finally, we provide a linear programming formulation for the exact computation in \Cref{section:ilp}.

\begin{figure}
	\begin{subfigure}{0.55\linewidth}\centering
		\scalebox{1}{\includegraphics[width=.9\linewidth]{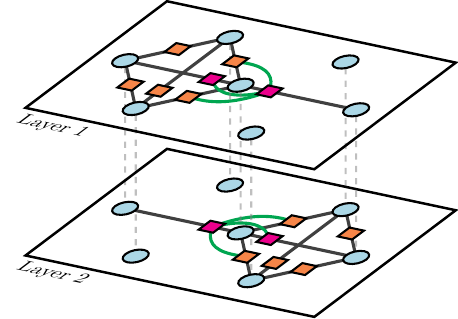}}  
		\caption{}
		\label{fig:example3:a}
	\end{subfigure}\hfill%
	\begin{subfigure}{0.425\linewidth}\centering
		\scalebox{1}{\includegraphics[width=.9\linewidth]{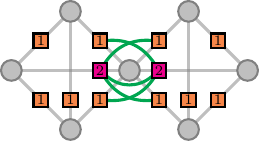}}
		\caption{}
		\label{fig:example3:b}
	\end{subfigure}\vspace{-.1cm}
	\caption{Construction of the combined wedge graph for \Cref{alg:klayerstc}. (a)~A multilayer graph with two layers. In each layer, the wedge graph is overlaid with its nodes shown as squares and its edges in green. (b)~The combined node-weighted wedge graph, with the aggregated multilayer graph shown in gray in the background.}
	\label{fig:example3}
\end{figure}

\section{Multilayer STC with Edge Insertions}\label{sec:mlstcplus}

We now consider the \textsc{MinMultiLayerSTC+} problem (\Cref{def:multistcplus}), where new \emph{weak} edges $E_i^+\subseteq {V \choose 2}\setminus E_i$ can be added to layer $i\in[k]$ to potentially achieve a better overall STC labeling. 
The problem formulation follows the same axiomatic principles introduced in \Cref{sec:mlstc} (Axioms 1-5) to govern the cost function.
The goal is to minimize the total number of \emph{weak} edges (original ones in $E_i \setminus S_i$ plus newly added ones in $E_i^+$) plus the disagreement cost $d_k(S_1,\ldots,S_k)$. The disagreement cost $d_k$ is now defined over edges in $(E_i\cup E_i^+)\cap (E_j\cup E_j^+)$.

\begin{problem}[\textsc{MinMultiLayerSTC+}]\label{def:multistcplus}
	Given a multilayer graph $G=(V, E_1, \ldots, E_k)$, find $k$ subsets of new edges $E_i^+\subseteq {V \choose 2}\setminus E_i$ and 
	labelings $S_i\subseteq E_i$ for $i\in[k]$ such that
	\begin{enumerate}
		\item for each $i\in [k]$, $S_i$ satisfies the STC for $E_i\cup E_i^+$, and 
		\item 
		$
		\sum_{i\in[k]} |E_i^+\cup E_i\setminus S_i| + d_k(S_1,\ldots,S_k) \text{~~is minimized.}
		$
	\end{enumerate}
	
\end{problem} 

\Cref{def:multistcplus} generalizes the single-layer variant of the problem defined by \cite{sintos2014using}.
Similar to \Cref{lemma:eqgeneral} for \textsc{MinMultiLayerSTC}, the following lemma establishes that an optimal solution to \textsc{MinMultiLayerSTC+} can be found that is consistent across layers, even when considering newly added edges.
\begin{lemma}\label{lemma:eqgeneralplus}
	Let $S_1^*, \ldots, S_k^*$ be labelings of strong edges for a multilayer graph $G=(V, E_1, \ldots, E_k)$ with potential new edges $E_i^+ \subseteq {V \choose 2} \setminus E_i$, where $S_i^* \subseteq E_i$ for each $i \in [k]$. Let $C_{+}^*$ be their objective value for the \textsc{MinMultiLayerSTC+} problem, i.e., 
	$C_{+}^* = \sum_{i\in[k]} |(E_i \cup E_i^+) \setminus S_i^*| + d_k(S_1^*,\ldots,S_k^*)$.
	There exists another set of labelings $H^*=\{H_1^*, \ldots, H_k^*\}$ (with $H_i^* \subseteq E_i$) s.t.:
	\begin{enumerate}
		\item The labeling $H^*$ is consistent, i.e., $d_k(H_1^*, \ldots, H_k^*) = 0$.
		\item The objective value of $H^*$ for \textsc{MinMultiLayerSTC+} is equal to $C_{+}^*$, i.e., $\sum_{i\in[k]} |(E_i \cup E_i^+) \setminus H_i^*| + d_k(H_1^*, \ldots, H_k^*) = C_{+}^*$.
	\end{enumerate}
\end{lemma}

\Cref{lemma:eqgeneralplus} implies that the cost of an optimal consistent solution involves summing, for each edge $e \in E \cup \bigcup E_i^+$, its contribution if it is \emph{weak}. If $e \in E_i$ (original), it contributes to $|E_i \setminus S_i|$. If $e \in E_i^+$ (new), it contributes to $|E_i^+|$. Due to consistency, if $e$ is \emph{weak}, it is weak in all layers it's part of (original or added).
To develop our approximation algorithm, we first introduce an intermediate problem variant.
\begin{problem}[\textsc{ConstraintMmlSTC+}]\label{def:consistentmultistcplus}
	As \textsc{MinMultiLayerSTC+} (\Cref{def:multistcplus}), with the additional constraint (3): for any $i\in[k]$, if a new edge $e \in E_i^+$ is added to close a wedge in $G_i=(V,E_i)$, and if for some other layer $j\neq i$, $e$  also closes a wedge in $G_j$ (i.e., $(v, e)\in \mathcal{W}(E_j)$ for some $v$), then $e \in E_j^+$ must also hold.
\end{problem} 
The new constraint (3) means that if a new edge $e$ is chosen to be added as \emph{weak} to resolve STC violations, its addition is propagated to all layers where it could serve a similar purpose. This aids in constructing a unified hypergraph for approximation. The relationship between this variant and our target problem is given by:

\begin{lemma}\label{lemma:doubleapprox}
	An $\alpha$-approximation for \textsc{ConstraintMmlSTC+} (\Cref{def:consistentmultistcplus}) provides a $2\alpha$-approximation for \textsc{MinMultiLayerSTC+} (\Cref{def:multistcplus}).
\end{lemma}

Our approximation algorithm, \Cref{alg:klayerstcplus}, addresses \textsc{MinMultiLayerSTC+} by first targeting \textsc{ConstraintMmlSTC+}.
It constructs a combined 3-uniform wedge hypergraph $W=(V_W, E_W)$. The vertices $V_W$ correspond to nodes $n_e$ for all existing edges $e \in E$ (the aggregated set of unique edges) and all potential new edges $e' \in {V \choose 2} \setminus E$.
Hyperedges in $E_W$ represent open wedges from any layer $G_i=(V,E_i)$. 
The weight $\omega(n_e)$ for a node $n_e \in V_W$ is:
\begin{itemize}[nosep,leftmargin=1.5em]
	\item If $e \in E$: $\omega(n_e)$ is its multiplicity, $|\{j \mid e \in E_j\}|$.
	\item If $e \notin E$ (a potential new edge): $\omega(n_e)$ is the number of layers $j$ where adding $e$ as \emph{weak} would close an open wedge in $G_j$.
\end{itemize}
The algorithm then finds an approximate Minimum Weighted Vertex Cover (MWVC) $C \subseteq V_W$ on this hypergraph $W$ (e.g., using a 3-approximation algorithm like pricing for 3-uniform hypergraphs). Edges $e \in E$ with $n_e \in C$ are designated as \emph{weak}. Potential new edges $e' \notin E$ with $n_{e'} \in C$ are provisionally added to the sets $E_i^+$ for all layers $i$ where their addition was counted in $\omega(n_{e'})$.

\begin{algorithm}
	\caption{A 6-approximation algorithm for \textsc{MinMultiLayerSTC+}}\label{alg:klayerstcplus}
	\Input{Multilayer graph $G=(V,E_1,\ldots,E_k)$} 
	\Output{Labelings $S_1,\ldots,S_k$ and new edge sets $E_1^+, \dots, E_k^+$}
	\tcp{Construct 3-uniform wedge hypergraph $W=(V_W, E_W)$}
	$V_W \gets \{n_e \mid e \in E \text{ or $e\not\in E$ and closes at least one open wedge}\}$ \;
	$E_W \gets \emptyset$\;
	\For{$i \in \{1,\ldots,k\}$}{
		\For{each open wedge $(v, \{u,x\})$ in $G_i=(V,E_i)$}{
			$E_W \gets E_W \cup \{\{n_{vu}, n_{vx}, n_{ux}\}\}$\;
		}
	}
	\For{$n_e \in V_W$}{
		\If{$e \in E = \bigcup E_j$ }{
			$\omega(n_e) \gets |\{j \mid e \in E_j\}|$\;
		} \Else (\tcp*[h]{$e$ is potential new edge}){
			$\omega(n_e) \gets |\{j \mid \exists v: (v,\{u,w\})$ \text{ is wedge in } $G_j \text{ and } e=\{u,w\}  \}|$
		}
	}
	Approximate MWVC $C \subseteq V_W$ on $W$ with weights $\omega$\;
	
	$E_\text{weak} \gets \{e \mid n_e \in C \text{ and } e \in E\}$\;
	Initialize $E_i^+ \gets \emptyset$ for all $i \in [k]$\;
	\For{$n_e \in C$ where $e \notin E$ (new edge selected by MWVC)}{
		\For{$i \in [k]$ such that adding $e$ to $G_i$ closes an open wedge in $G_i$}{ 
			Add $e$ to $E_i^+$ 
		}
	}
	
	\For{$i\in[k]$}{
		$S_i\gets E_i\setminus E_\text{weak}$ \tcp*{Strong labeled edges}
	}
	\Return{strong edges $S_i$ and new edges $E_i^+$ for each $i\in [k]$}
\end{algorithm}

\begin{theorem}\label{theorem:approxstcplus}
	Given an $\alpha$-approximation for the MWVC in $3$-uniform hypergraphs with running time $T_{VC\_H}$,
	\Cref{alg:klayerstcplus} computes a $2\alpha$-approximation for \textsc{MinMultiLayerSTC+} (\Cref{def:multistcplus}) in $\mathcal{O}(k\cdot nm + T_{VC\_H})$ running time.
\end{theorem}

Our algorithm (\Cref{alg:klayerstcplus}) achieves a 6-approximation for the \textsc{MinMultiLayerSTC+} problem by employing the pricing method, a linear-time 3-approximation algorithm for MWVC in 3-uniform hypergraphs~\citep{bar1981linear}. The overall time and space complexity is $\mathcal{O}(k \cdot nm)$. For completeness, we also provide an ILP formulation for the exact computation in \Cref{section:ilp}.

\section{Experiments}\label{sec:experiments}

\begin{table}
	\centering
	\caption{Statistics of the real-world datasets. }  
	\label{table:datasets_stats}
		\renewcommand{\arraystretch}{0.9} \setlength{\tabcolsep}{8pt}
		\begin{tabular}{lrrrrr}\toprule
			\textbf{Dataset}        &   $|V(G)|$      & $|\bigcup_i E_i(G)|$ & $\sum_i| E_i(G)|$ &   \#Layers & Domain 
			\\ \midrule
			
			\emph{AUCS}      &       61 &         353 &    620 &  5 &  Social network \\
			\emph{Hospital}  &       75 &     1\,139  & 1\,885 &  5 &  Face-to-face \\
			\emph{Airports}  &      417 &     2\,953  & 3\,588 & 37 &  Transportation\\
			\emph{Rattus}    &   2\,634 &     3\,677  & 3\,905 &  6 &  Biological \\
			\emph{FfTwYt}    &   6\,401 &     60\,583 & 74\,810 &  3 &  Social network \\
			\emph{Knowledge} & 14\,505  &    210\,946 & 225\,898 & 30 &  Knowledge graph \\
			\emph{HomoSap}   &  18\,190 &    137\,659 & 153\,922 &  7 &  Biological \\	
			\emph{DBLP}		 & 344\,814 & 1\,528\,399 & 1\,837\,562 & 168  &  Collaboration \\
			\bottomrule
		\end{tabular}
\end{table}

We conduct experimental evaluations to answer the following research questions:
\begin{itemize}
	\item[\textbf{Q1.}] \textbf{Consistency:} How much does the multilayer STC improve the consistency compared to the standard STC?
	\item[\textbf{Q2.}] \textbf{Multilayer STC+:} How does the \textsc{MinMultiLayerSTC+} improve the labeling compared to \textsc{MinMultiLayerSTC}? 
	\item[\textbf{Q3.}] \textbf{Approximation quality:} How is the empirical approximation quality of our algorithms?
	\item[\textbf{Q4.}] \textbf{Efficiency:} How efficient are our algorithms in terms of running time and memory usage?
\end{itemize}
Additionally, we present a qualitative case study in \Cref{sec:use_case}.

\subsection{Algorithms, Datasets, and Metrics}
We implemented and compared the following three groups of algorithms:

\begin{itemize}
	\item \textbf{Ours:} \alg and \algp are our approximation algorithms for the multilayer STC and STC+ problems (\Cref{alg:klayerstc} and \Cref{alg:klayerstcplus}). \texttt{ExactML} and \texttt{ExactML+} are the exact computations of the STC and STC+ using the ILPs introduced in \Cref{section:ilp}. 
	\item \textbf{Layer-wise baselines:} \texttt{BLExact} and \texttt{BLExact+} are exact baselines that apply the single-layer ILP formulations for STC~\citep{adriaens2020relaxing} and STC+~\citep{veldt2022correlation}, respectively, independently to each layer.  Similarly, \texttt{BLstc} and \texttt{BLstc+} are baseline heuristics that approximate the solutions of STC and STC+, respectively, in each layer independently. 
	\item \revision{\textbf{Aggregation baselines:} \texttt{BLAgg} and \texttt{BLAgg+} are baselines that aggregate layers via edge multiplicity and then solve a single weighted STC or STC+ instance on the resulting graph.
	The resulting labels are mapped back to all layers in which an edge appears.
	In addition, we report exact variants of the aggregated baselines (\texttt{BLAggExact}, \texttt{BLAggExact+}), obtained by solving the corresponding weighted STC/STC+ instance via ILP.}
\end{itemize}

For the approximation-based methods, we compute the required vertex covers in the wedge graphs using the pricing algorithm (with unit weights in the unweighted case)~\citep{bar1981linear}
and a greedy algorithm that iteratively adds the vertex covering the largest number of uncovered edges (see~\cite{vazirani2001approximation}).
We append \texttt{(P)} for pricing and \texttt{(G)} for greedy to the algorithm names, e.g., \algvp for our STC approximation using the pricing method.

\revision{The aggregation-based baselines capture a common heuristic in multilayer analysis: reducing the network to a single weighted graph before applying a single-layer method. This enforces identical labels across layers for shared edges and thus eliminates cross-layer conflicts. However, since STC constraints are evaluated independently in each layer, aggregation can mask layer-specific open wedges and may produce solutions that violate STC in individual layers. We therefore additionally report the number of STC violations per layer for these baselines.}

\revision{Finally, we do not compare against learning-based tie-strength prediction methods. These are typically studied in supervised, single-layer settings under application-specific definitions of tie strength (see~\cite{cheng2025bts}), whereas our work addresses an unsupervised structural optimization problem with explicit multilayer STC constraints. We therefore focus on optimization-based baselines targeting the same problem.}

We implemented our algorithms in C++ using GNU CC Compiler~11.4.0 with the flag \texttt{--O3}. We used Gurobi 11 for solving ILPs.
All experiments ran on a computer cluster. Each experiment had an exclusive node with an Intel(R) Xeon(R) Gold 6130
CPU @ 2.10GHz and 96~GB of RAM. We used a time limit of 12 hours.
The source code and all datasets are available at \url{https://gitlab.com/multilayergraphs/multilayerstc}.

\smallskip
\noindent
\textbf{Datasets.} We use eight real-world datasets from different domains. \Cref{table:datasets_stats} provides an overview and \Cref{appendix:datasets} further details.

\smallskip
\noindent
\revision{\textbf{Metrics.}
To assess cross-layer coherence, we evaluate two complementary measures:
\begin{enumerate}
	\item \textbf{Edge disagreement rate.}
	We define
	\[
	\ecr(S)
	= 1-\frac{1}{|F|}\sum_{e\in F}\frac{s(e)}{s(e)+w(e)},
	\quad
	F=\{e\in E : s(e)>0\}.
	\]
	This measures the average cross-layer inconsistency of edges
	that are labeled strong in at least one layer.
	It equals $0$ when all such edges are consistently strong
	and increases up to $1$ as strong labels become fragmented across layers.
	\item \textbf{Weak-edge disagreement rate.}
	We define
	\[
	\wecr(S)
	= \frac{|\{e \in E : s(e)>0 \land w(e)>0\}|}
	{|\{e \in E : w(e)>0\}|},
	\]
	which measures the fraction of weak-labeled edges that are also labeled strong in
	at least one other layer, i.e., how frequently conflicts occur among edges that are
	classified as weak.
\end{enumerate}
In addition, we report the fraction of weak edges, running times, and where applicable, the number of STC violations for aggregation baselines that do not enforce layer-wise feasibility. An overview of the reported measures per method type is given in \Cref{tab:reported_measures}.}

\begin{table*}[ht]
	\centering
	\caption{\revision{Reported measures. A \checkmark indicates that the metric is computed and reported; otherwise, the metric is 0 by construction.} 
	}
	\label{tab:reported_measures}
	\resizebox{1\linewidth}{!}{%
	{\color{black} 
	\renewcommand{\arrayrulecolor}{blue} 
	\begin{tabular}{lccccc}
		\toprule
		\textbf{Method type} & Weak frac. & $\ecr(S)$ & $\wecr(S)$ & STC viol. & Time \\
		\midrule
		\textbf{Layer-wise baselines} (\texttt{BLexact}, \texttt{BLexact+}, \texttt{BLstc}, \texttt{BLstc+}) & \checkmark & \checkmark & \checkmark & 0 & \checkmark \\
		\textbf{Aggregation baselines} (\texttt{BLAggExact}, \texttt{BLAggExact+}, \texttt{BLAgg}, \texttt{BLAgg+}) & \checkmark & 0 & 0  & \checkmark & \checkmark \\
		\textbf{Ours} (\texttt{ExactML}, \alg, \algp) & \checkmark & 0 &  0 & 0 & \checkmark \\
		\bottomrule
	\end{tabular}}}
\end{table*}

\begin{table*}[!htbp]
\caption{\revision{Results for multilayer STC and STC+ (lower is better for all metrics).
	Our algorithms enforce consistency by construction and satisfy all STC constraints.}}

\begin{subtable}[b]{1\textwidth}
	\centering
	\caption{Multilayer STC results.}
	\label{table:stc_all_rotated}
	\renewcommand{\arraystretch}{1}
	\resizebox{1\linewidth}{!}{%
		\begin{tabular}{lllcccccccc} 
			\toprule
			& \textbf{Method} & \textbf{Metric} & \emph{AUCS} & \emph{Hospital} & \emph{Airports} & \emph{Rattus}
			& \emph{FfTwYt} & \emph{Knowledge} & \emph{HomoSap} & \emph{DBLP} \\
			\midrule
			
			\multirow{9}{*}{\rotatebox{90}{\textbf{Layer-wise bl.}}} 
			& \multirow{3}{*}{\texttt{BLExact}}
			& Weak& 0.55 & 0.75 & 0.85 & 0.78 & -- & -- & -- & -- \\
			&& $\ecr(S)$  & 0.16 & 0.25 & 0.08 & 0.06 & -- & -- & -- & -- \\
			&& $\wecr(S)$  & 0.23 & 0.16 & 0.03 & 0.03 & -- & -- & -- & -- \\
			\cmidrule(lr){2-11}
			
			& \multirow{3}{*}{\texttt{BLstc(P)}}
			& Weak & 0.94 & 0.98 & 0.95 & 0.90 & 0.99 & 0.97 & 0.97 & 0.64 \\
			&& $\ecr(S)$  & 0.59 & 0.65 & 0.90 & 0.86 & 0.68 & 0.89 & 0.83 & 0.90 \\
			&& $\wecr(S)$ & 0.25 & 0.12 & 0.02 & 0.03 & 0.01 & 0.01 & 0.01 & 0.11 \\
			\cmidrule(lr){2-11}
			
			& \multirow{3}{*}{\texttt{BLstc(G)}}
			& Weak & 0.91 & 0.95 & 0.91 & 0.84 & 0.96 & 0.93 & 0.94 & 0.55 \\
			&& $\ecr(S)$  & 0.44 & 0.58 & 0.87 & 0.88 & 0.69 & 0.90 & 0.84 & 0.89 \\
			&& $\wecr(S)$ & 0.32 & 0.19 & 0.04 & 0.03 & 0.03 & 0.01 & 0.02 & 0.15 \\
			\midrule
			
			\multirow{6}{*}{\rotatebox{90}{\textbf{Aggregated bl.}}} 
			& \multirow{2}{*}{\texttt{BLAggExact}}
			& Weak& 0.52 & 0.72 & 0.82 & 0.81 & -- & -- & -- & -- \\
			&& STC viol. & 140 & 824 & 2\,107 & 31 & -- & -- & -- & -- \\
			\cmidrule(lr){2-11}
			
			& \multirow{2}{*}{\texttt{BLAgg(P)}}
			& Weak& 0.73 & 0.89 & 0.96 & 0.90 & 0.98 & 0.98 & 0.98 & 0.77 \\
			&& STC viol. & 110 & 332 & 118 & 7 & 122 & 569 & 621 & 4\,732 \\
			\cmidrule(lr){2-11}
			
			& \multirow{2}{*}{\texttt{BLAgg(G)}}
			& Weak& 0.64 & 0.83 & 0.88 & 0.84 & 0.94 & 0.94 & 0.94 & 0.66 \\
			&& STC viol. & 161 & 794 & 2\,031 & 19 & 1\,555 & 4\,043 & 4\,337 & 26\,975 \\
			\midrule
			
			\multirow{3}{*}{\rotatebox{90}{\textbf{Ours}}} 
			& \multirow{1}{*}{\texttt{ExactML}}
			& Weak& 0.65 & 0.84 & 0.87 & 0.79 & -- & -- & -- & -- \\
			& \multirow{1}{*}{\algvp}
			& Weak& 0.81 & 0.96 & 0.94 & 0.88 & 0.97 & 0.96 & 0.97 & 0.58 \\
			& \multirow{1}{*}{\algvg}
			& Weak& 0.67 & 0.85 & 0.88 & 0.79 & 0.93 & 0.92 & 0.92 & 0.47 \\
			\bottomrule
	\end{tabular}}
\end{subtable}

\begin{subtable}[b]{1\textwidth}
	\centering
	\caption{Multilayer STC+ results.}
	\label{table:stcplus_all_rotated}
	\renewcommand{\arraystretch}{0.95}
	\setlength{\tabcolsep}{6pt}
	\resizebox{1\linewidth}{!}{%
		\begin{tabular}{lllcccccccc}
			\toprule
			& \textbf{Method} &\textbf{Metric}
			& \emph{AUCS} & \emph{Hospital} & \emph{Airports} & \emph{Rattus}
			& \emph{FfTwYt} & \emph{Knowledge} & \emph{HomoSap} & \emph{DBLP} \\
			\midrule
			
			\multirow{9}{*}{\rotatebox{90}{\textbf{Layer-wise bl.}}}
			& \multirow{3}{*}{\texttt{BLExact+}}
			& Weak     & 0.45 & 0.54 & 0.80 & ~0.75* & -- & -- & -- & -- \\
			&	& $\ecr(S)$  & 0.12 & 0.19 & 0.14 & ~0.05* & -- & -- & -- & -- \\
			&	& $\wecr(S)$  & 0.23 & 0.23 & 0.04 & 0.03 & -- & -- & -- & -- \\
			\cmidrule(lr){2-11}
			
			&	\multirow{3}{*}{\texttt{BLstc+(P)}}
			& Weak& 0.95 & 0.96 & 0.96 & 0.90 & 0.99 & 0.97 & 0.98 & 0.64 \\
			&	& $\ecr(S)$  & 0.54 & 0.61 & 0.89 & 0.86 & 0.69 & 0.90 & 0.83 & 0.90 \\
			&	& $\wecr(S)$  & 0.21 & 0.08 & 0.02 & 0.03 & 0.01 & 0.01 & 0.01 & 0.10 \\
			\cmidrule(lr){2-11}
			
			&	\multirow{3}{*}{\texttt{BLstc+(G)}}
			& Weak& 0.88 & 0.95 & 0.83 & 0.79 & 0.90 & 0.89 & 0.92 & 0.54 \\
			&	& $\ecr(S)$  & 0.39 & 0.58 & 0.84 & 0.86 & 0.67 & 0.90 & 0.84 & 0.89 \\
			&	& $\wecr(S)$  & 0.36 & 0.28 & 0.07 & 0.03 & 0.04 & 0.02 & 0.03 & 0.15 \\
			\midrule
			
			\multirow{6}{*}{\rotatebox{90}{\textbf{Aggregated bl.}}}	
			& \multirow{2}{*}{\texttt{BLAggExact+}}
			& Weak& 0.34 & 0.24 & 0.68 & 0.77 & -- & -- & -- & -- \\
			&	& STC viol. & 583 & 10\,550 & 7\,807 & 224 & -- & -- & -- & -- \\
			\cmidrule(lr){2-11}
			
			&	\multirow{2}{*}{\texttt{BLAgg+(P)}}
			& Weak& 0.72 & 0.91 & 0.97 & 0.90 & 0.98 & 0.98 & 0.98 & 0.77 \\
			&	& STC viol. & 102 & 278 & 35 & 18 & 142 & 556 & 537 & 5\,293 \\
			\cmidrule(lr){2-11}
			
			&	\multirow{2}{*}{\texttt{BLAgg+(G)}}
			& Weak& 0.46 & 0.49 & 0.78 & 0.78 & 0.91 & 0.91 & 0.90 & 0.64 \\
			&	& STC viol. & 466 & 6\,205 & 5\,629 & 226 & 6\,137 & 22\,831 & 37\,966 & 42\,271 \\
			
			\midrule
			\multirow{3}{*}{\rotatebox{90}{\textbf{Ours}}} 
			&	\multirow{1}{*}{\texttt{ExactML+}}
			& Weak& 0.58 & ~0.76* & ~0.84* & ~0.77* & -- & -- & -- & -- \\
			&	\multirow{1}{*}{\algpvp}
			& Weak& 0.80 & 0.92 & 0.91 & 0.87 & 0.97 & 0.95 & 0.96 & 0.58 \\
			&	\multirow{1}{*}{\algpvg}
			& Weak& 0.64 & 0.83 & 0.81 & 0.74 & 0.92 & 0.90 & 0.90 & 0.46 \\
			\bottomrule
	\end{tabular}}
\end{subtable}
\end{table*}

\subsection{Results}
We demonstrate that our methods achieve perfect consistency while maintaining competitive solution quality, strong approximation guarantees, and practical efficiency.

\smallskip
\textbf{Q1.~Consistency.} \revision{
\Cref{table:stc_all_rotated} reveals fundamental limitations of layer-wise approaches.
The exact baseline (\texttt{BLExact}) exhibits $\ecr(S) \in [0.05, 0.25]$ across all datasets, confirming that independent layer optimization systematically produces conflicting tie semantics.
Approximation baselines amplify this issue dramatically: \texttt{BLstc(P)} and \texttt{BLstc(G)} yield $\ecr(S) > 0.8$ on six of eight datasets, with values reaching 0.90 (\emph{DBLP}, \emph{Airports}, \emph{Knowledge}), indicating that 90\% of strong edges have inconsistent labels across layers.
Moreover, $\wecr(S)$ values up to 0.32 show that conflicts contaminate substantial fractions of weak edges, undermining interpretability.}

\revision{Aggregation baselines fare worse: while producing lower weak fractions by exploiting cross-layer evidence, they violate layer-specific STC constraints extensively (up to 26\,975 violations on \emph{DBLP}, 10\,550 on \emph{Hospital} for STC+), rendering solutions invalid.}

\revision{In contrast, our methods enforce $\ecr(S) = \wecr(S) = 0$ by construction with zero STC violations, while achieving weak fractions competitive with or superior to layer-wise baselines. Critically, $\algvg$ consistently outperforms $\texttt{BLstc(G)}$ in weak fraction (e.g., 0.47 vs. 0.55 on \emph{DBLP}), demonstrating that consistency does not require sacrificing solution quality.}

\begin{figure}
	\centering
	\begin{subfigure}{0.5\linewidth}
		\centering
		\includegraphics[width=0.8\linewidth]{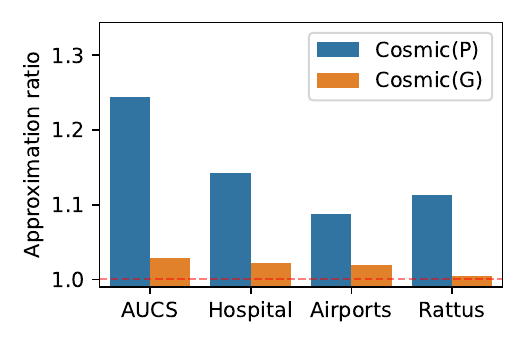} 
		\caption{STC}
		\label{fig:approxqualitya}
	\end{subfigure}\hfill%
	\begin{subfigure}{0.5\linewidth}
		\centering
		\includegraphics[width=0.8\linewidth]{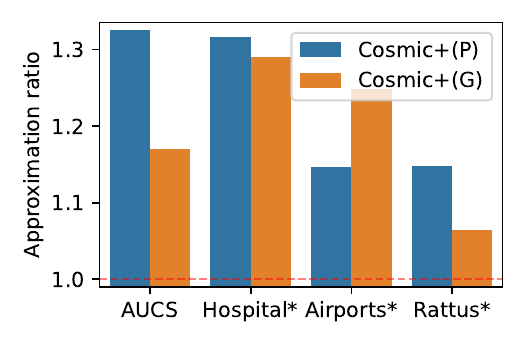}
		\caption{STC+}
		\label{fig:approxqualityb}
	\end{subfigure}
	\caption{The empirical approximation ratios. For datasets with (*), the reported ratio is an upper bound on the true empirical approximation ratio, computed using the best-known lower bound on the optimal solution found within the 12h time limit.}
	\vspace{-.3cm} 
	\label{fig:approxquality}
\end{figure}

\begin{table*}[htb]
	\centering
	\caption{Running times in seconds (s).}
	\label{table:runningtimes}
	\renewcommand{\arraystretch}{1.1}
	\setlength{\tabcolsep}{6pt}
	\sisetup{table-number-alignment=center}
	\resizebox{1\linewidth}{!}{%
		\begin{tabular}{c l
				S[table-format=3.3] S[table-format=3.2] S[table-format=3.2] S[table-format=3.2]
				S[table-format=3.2] S[table-format=3.2] S[table-format=3.2] S[table-format=3.2]}
			\toprule
			& \textbf{Method}
			& {\emph{AUCS}} & {\emph{Hospital}} & {\emph{Airports}} & {\emph{Rattus}}
			& {\emph{FfTwYt}} & {\emph{Knowledge}} & {\emph{HomoSap}} & {\emph{DBLP}} \\
			\midrule
			
			\multirow{6}{*}{\rotatebox{90}{\textsc{STC}}}
			& \texttt{BLstc(P)}  & 0.005 & 0.01 & 0.01 & 0.12 & 7.94  & 22.99 & 41.20  & 69.59 \\
			& \texttt{BLstc(G)}  & 0.005 & 0.01 & 0.01 & 0.11 & 9.74  & 26.56 & 45.78  & 94.01 \\
			& \texttt{BLAgg(P)}  & 0.001 & 0.01 & 0.03 & 0.10 & 4.70  & 26.40 & 27.01 & 18.79 \\
			& \texttt{BLAgg(G)}  & 0.001 & 0.01 & 0.04 & 0.15 & 11.37 & 70.34 & 66.72 & 71.00 \\
			& \algvp             & 0.005 & 0.01 & 0.03 & 0.21 & 9.35  & 32.82 & 54.75  & 12.85 \\
			& \algvg             & 0.005 & 0.01 & 0.04 & 0.31 & 16.83 & 58.91 & 100.04 & 25.87 \\
			\midrule
			
			\multirow{6}{*}{\rotatebox{90}{\textsc{STC+}}}
			& \texttt{BLstc+(P)} & 0.003 & 0.01 & 0.04 & 0.59 & 29.45  & 90.87  & 231.46 & 59.03 \\
			& \texttt{BLstc+(G)} & 0.005 & 0.03 & 0.10 & 1.51 & 103.20 & 340.73 & 905.07 & 127.38 \\
			& \texttt{BLAgg+(P)} & 0.003 & 0.02 & 0.10 & 0.70 & 31.87  & 199.61 & 242.05 & 184.75 \\
			& \texttt{BLAgg+(G)} & 0.005 & 0.04 & 0.27 & 1.68 & 106.12 & 673.57 & 775.10 & 512.68 \\
			& \algpvp            & 0.003 & 0.01 & 0.07 & 0.67 & 42.52  & 102.82 & 272.38 & 50.21 \\
			& \algpvg            & 0.005 & 0.03 & 0.16 & 1.57 & 118.27 & 366.63 & 731.67 & 115.28 \\
			\bottomrule
	\end{tabular}}
\end{table*}

\begin{table}
	\centering
	\caption{Sizes of the wedge (hyper-)graphs (the number of (hyper-)edges is equal for STC and STC+). }  
	\label{table:wedgegraphsizes}
		\renewcommand{\arraystretch}{1} \setlength{\tabcolsep}{11pt}
		\begin{tabular}{lrrr}
			\toprule
			
			\textbf{Dataset}& (STC) $|V(W)|$ & (STC+) $|V(W)|$ & $|E(W)|$ \\ \midrule
			\emph{AUCS}      &         353 &          845 &       2\,152 \\
			\emph{Hospital}  &      1\,139 &       2\,333 &      15\,196 \\
			\emph{Airports}  &      2\,953 &      17\,053 &      50\,226 \\
			\emph{Rattus}    &      3\,677 &     380\,818 &     386\,877 \\
			\emph{FfTwYt}    &     60\,583 &  6\,410\,083 & 12\,037\,201 \\
			\emph{Knowledge} &    210\,946 & 19\,795\,949 & 42\,820\,276 \\
			\emph{HomoSap}   &    137\,659 & 50\,630\,405 & 62\,634\,802 \\
			\emph{DBLP}      & 1\,528\,399 &  7\,925\,839 & 12\,356\,011 \\ 
			\bottomrule
		\end{tabular}
	\end{table}

\smallskip
\textbf{Q2.~Multilayer STC+.} \revision{
\Cref{table:stcplus_all_rotated} confirms that edge insertion systematically improves solution quality while preserving perfect consistency. 
Comparing Tables~\ref{table:stc_all_rotated} and \ref{table:stcplus_all_rotated}, STC+ reduces weak fractions substantially: \texttt{ExactML+} achieves 0.58 vs. 0.65 (\emph{AUCS}), 0.76 vs. 0.84 (\emph{Hospital}), and 0.84 vs. 0.87 (\emph{Airports}) compared to \texttt{ExactML}.
Our approximations maintain this advantage across all datasets and $\algpvg$ reduces weak fractions by 3-8 percentage points versus $\algvg$.}

\revision{Layer-wise baselines again exhibit severe consistency violations ($\ecr(S)$ up to 0.90), while aggregation baselines produce massive STC violations (up to 42\,271 on \emph{DBLP}).
The pattern is clear: allowing edge insertions provides flexibility to reduce weak assignments, but only our formulation ensures this flexibility is exercised consistently across layers while maintaining STC validity everywhere.}

\smallskip
\textbf{Q3.~Approximation Quality.}
\Cref{fig:approxqualitya} shows the empirical approximation ratios, i.e., the approximated objective value divided by the optimal objective value computed by the ILPs for the multilayer STC and the four smaller datasets. 
Even though \Cref{alg:klayerstc} using the pricing method (\algvp) is theoretically a 2-approximation, the empirical approximation quality is much better with values between 1.24 for \emph{AUCS} and 1.08 for \emph{Airports}.
The greedy variant \algvg, even though it is an $\mathcal{O}(\log n)$ approximation, has even better empirical approximation ratios between 1.005 for \emph{Rattus} and 1.03 for \emph{AUCS}. This is expected, as the greedy vertex cover algorithm is known to often outperform the pricing method~\citep{gomes2006experimental}.

\Cref{fig:approxqualityb} shows the results for the STC+. Here, we use the best lower bounds for the optimal solutions for the datasets for which no optimal solution could be obtained within the time limit, and hence, the corresponding approximation ratios reported are upper bounds.
We see low empirical approximation ratios for both \algpvp and \algpvg with values between 1.14 for \emph{Airports} and 1.32 for \emph{AUCS} in case of \algpvp and 1.06 for \emph{Rattus} and 1.29 for \emph{Hospital} in the case of \algpvg. Here, \algpvp beats \algpvg for the \emph{Airports} dataset.

\smallskip
\textbf{Q4.~Efficiency.}
\Cref{table:runningtimes} shows that running times scale with wedge (hyper-)graph sizes (cf.~\Cref{table:wedgegraphsizes}).
On smaller networks (\emph{AUCS}, \emph{Hospital}), all methods complete in $<0.1$s.
On larger networks, our algorithms remain competitive: $\algvp$ processes \emph{DBLP} (168 layers, 1.8M edges) in 13s versus 70s for \texttt{BLstc(P)}, despite solving a harder problem.
The advantage is structural: we solve a single multilayer wedge graph, while layer-wise baselines invoke vertex cover once per layer (168 times for \emph{DBLP}), which becomes costly as the number of layers grows.
Aggregation baselines are often competitive in runtime since they operate on a single aggregated graph, but this simplification ignores layer-specific wedge constraints, explaining their substantial STC violations.
For STC+, similar patterns hold. In particular, $\algpvg$ outperforms \texttt{BLstc+(G)} on \emph{HomoSap}, as the baseline processes each layer separately while our method works on one multilayer wedge hypergraph.
Memory usage is dominated by wedge (hyper-)graph size, peaking on \emph{HomoSap} at 9.81\,GB (STC) and 18.64\,GB (STC+).\looseness=-1

\begin{figure}
	\centering
	\includegraphics[width=0.8\linewidth]{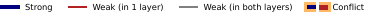}
	\begin{subfigure}{.9\linewidth}\centering
		\includegraphics[width=.9\linewidth]{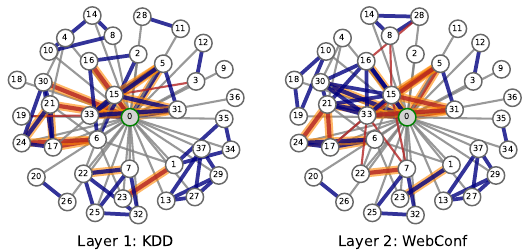}
		\caption{Inconsistent labeling computed with \texttt{BLExact}.}
		\label{fig:usecase:a}
	\end{subfigure}
	
	\begin{subfigure}{0.9\linewidth}\centering
		\includegraphics[width=.9\linewidth]{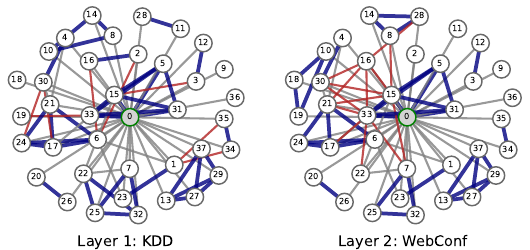}
		\caption{Consistent labeling computed with our \texttt{ExactML}.}
		\label{fig:usecase:b}
	\end{subfigure}
	\fbox{\resizebox{0.9\linewidth}{!}{%
			\begin{tabular}{lllll}
				0: C.~Zhang & 8: D.~Koutra & 16: K.~Shin & 24: S.~Yang & 32: X.~Song \\
				1: A.~Beutel & 9: D.~Eswaran & 17: M.~Jiang & 25: S.~Adeshina & 33: X.~Dong \\
				2: B.~Prakash & 10: D.~Chau & 18: M.~Lee & 26: S.~Günnemann & 34: Y.~Matsubara \\
				3: B.~Hooi & 11: E.~Papalexakis & 19: N.~Park & 27: T.~Januschowski & 35: Y.~Sakurai \\
				4: C.~Liu & 12: F.~Guo & 20: N.~Günnemann & 28: U.~Kang & 36: Y.~Yamaguchi \\
				5: C.~Zang & 13: J.~Gasthaus & 21: P.~Cui & 29: V.~Flunkert & 37: Y.~Wang \\
				6: C.~Faloutsos & 14: J.~Lee & 22: Q.~Zhu & 30: W.~Zhu &  \\
				7: D.~Zheng & 15: J.~Han & 23: R.~Rosenfeld & 31: X.~Li &  \\
			\end{tabular}
			
	}}\smallskip
	\caption{Labelings in an ego-network of \emph{Christos Faloutsos} (node 0 in the center) in a coauthorship multilayer network. The layers correspond to \emph{KDD} and \emph{WebConf}.
	}
	\label{fig:usecase}
\end{figure}

\subsection{Case Study: Consistent Tie Strength in DBLP}\label{sec:use_case}
We extracted a 2-layer ego network (of illustrative size) of \emph{Christos Faloutsos} from the DBLP co-authorship graph, using publications from KDD and WebConf to define the two layers. \Cref{fig:usecase} shows the results of computing STC separately for each layer and using our multilayer model. \emph{Strong} ties are shown in thick blue; \emph{weak} ties that appear in both layers in gray and otherwise in thinner red. Edges with disagreeing labels in the two layers are highlighted with an orange halo.

\revision{
The case study highlights a concrete issue caused by conflicting tie-strength labels. As shown in \Cref{fig:usecase:a}, solving STC independently in each layer can assign different strengths to the same coauthorship edge, labeling it \emph{strong} in one venue layer and \emph{weak} in another. In this ego network of Christos Faloutsos, 30 out of 175 total edges ($\approx 17\%$) are in disagreement. This inconsistency has immediate practical consequences. Consider a collaborator of Faloutsos whose coauthorship edge appears in both layers but is labeled \emph{strong} in KDD and \emph{weak} in WebConf. A strong label carries a concrete structural obligation under STC: if Faloutsos is strongly connected to this collaborator, then any other node to whom Faloutsos is also strongly connected must be at least weakly linked to that collaborator within the same layer. Labeling the same edge \emph{weak} in WebConf removes this obligation entirely in that layer, even if the underlying relationship has not changed. The edge therefore plays two structurally incompatible roles simultaneously: it enforces triangle closure in KDD but imposes no such constraint in WebConf.
Consequently, whether a given tie acts as a closure-enforcing strong relationship or a peripheral weak one is no longer a property of the relationship itself, but an artifact of which layer is consulted. This means that tasks relying on a single, coherent notion of tie strength, such as identifying Faloutsos's core collaborators, recommending new collaborations, or explaining which ties enforce closure constraints in the co-authorship network, yield different and incompatible answers depending on which layer is consulted. This issue is distinct from edges being present in one layer and absent in another, which merely reflects differences in publication activity across venues. Instead, the ambiguity arises when an edge is present in multiple layers but receives incompatible labels. The multilayer STC model (\Cref{fig:usecase:b}) eliminates these conflicts by enforcing a uniform label for each shared edge, producing a single tie-strength assignment that is valid across all layers in which the edge appears. 
}

\section{Conclusion and Future Work}
We introduced the first principled framework for inferring consistent tie strengths in multilayer networks. Our approach is grounded in a set of five natural axioms that lead to our objective function, ensuring that the resulting tie strength labels are both consistent and structurally well-justified. Our contributions include efficient approximation algorithms with constant-factor guarantees (2-approx. for STC, 6-approx. for STC+) and ILP formulations for the exact solution. Experiments on real-world networks confirm our methods yield consistent, interpretable labelings, outperforming baselines.\looseness=-1

\revision{Our framework is designed for semantically aligned multilayer networks where layers provide complementary views of the same underlying relationships.
	The explicit wedge enumeration ($O(knm)$ complexity) presents scalability challenges for web-scale graphs. Therefore, future work includes parallelization, sparsification, and sampling-based approximations to handle massive networks, and relaxed or learned consistency models for heterogeneous layers with different tie semantics.\looseness=-1}



\appendix

\section*{Appendix}

\begin{table}[htb!]
	\caption{Commonly used notations}
	\label{table:notation}
	\centering%
		\renewcommand{\arraystretch}{1.1}%
		\begin{tabular}{c@{\hspace{3mm}}l@{}}%
			\toprule
			\textbf{Symbol} & \textbf{Definition}
			\\\midrule
			$G=(V, E_1, \ldots, E_k)$ & multilayer graph \\
			$V$ & finite set of nodes \\
			$k$ & number of layers \\
			$E_1,\ldots,E_k$ & finite sets $E_i$ with $i\in[k]$ of undirected edges\\
			$E=\bigcup_{i\in[k]}E_i$ & aggregated set of edges in $G$ \\
			$n = |V|$ & number of nodes \\
			$m_i = |E_i|$ & number of edges in layer $i$\\
			$m = \sum_{i\in [k]} m_i$ & total number of edges in layers\\
			$(v,\{u,w\})$ & wedge or open triangle with edge $\{u,w\}$ missing\\
			$\mathcal{W}(E)$ & set of wedges in the set of edges $E$\\
			$\mathcal{W}(E,\{u,w\})$ & set of wedges in $E$ w.r.t.~$\{u,w\}$\\
			$W(G)$ & wedge graph of $G$\\
			$\omega(n_e)$ & weight of node $n_e$ in wedge (hyper-)graph\\
			$S_i \subseteq E_i$ & subset of \emph{strong} labeled edges in $E_i$ (labeling)\\
			$s(e)$ & number of layers in which $e$ is labeled \emph{strong}\\
			$w(e)$ & number of layers in which $e$ is labeled \emph{weak}\\
			$d_k(S_1,\ldots,S_k)$ & number of disagreements between labelings $S_1,\ldots, S_k$\\
			$\ecr(S)$ & Edge disagreement rate under labeling $S$\\
			$\wecr(S)$ & Weak-edge disagreement rate under labeling $S$\\
			\bottomrule
		\end{tabular}
\end{table}

\section{Omitted Proofs}

\begin{proof}[Proof of \Cref{theorem:costfunction}]
	$\Rightarrow:$ By Axioms (2)-(3), the total cost is a sum over edges and each edge's contribution depends only on its counts of strong and weak labels; hence
	$\mathrm{Cost}=\sum_{e\in E} f(s(e),w(e))$ for some $f$.
	
	By Axiom (1) with unit weak cost, each weak instance contributes $1$, so write
	$f(s(e),w(e))=w(e)+h(s(e),w(e))$ with $h(s(e),w(e))\ge 0$.  
	If $w(e)=0$, there is no strong/weak conflict, so by Axiom (4) there is no disagreement penalty; hence, (ii) follows with $h(s(e),0)=0$, i.e., $f(s(e),0)=0$.
	
	By Axiom (5), duplicating a weak-only layer must not change the
	\emph{disagreement} part; thus $h(s(e),w(e))$ cannot vary with the exact value of $w(e)$ once $w(e)>0$.
	Hence there exists $g$ with $h(s(e),w(e))=g(s(e))$ for all $w(e)>0$, giving
	$f(s(e),w(e))=w(e)+g(s(e))$ when $w(e)>0$ showing (iii). 
	Axiom (4) further requires that disagreement does not get cheaper as more layers label the edge strong, so $g$ is non-decreasing; also $g(0)=0$ (no conflict when $s(e)=0$).
	Finally, since $h\ge 0$, (i) follows as we have $f(s(e),w(e))\ge w(e)$.

	\noindent
	$\Leftarrow:$ 
	Assume there exists a non-decreasing function $g:\mathbb{N}_0\to\mathbb{R}_{\ge 0}$ with $g(0)=0$ and define
	\[
	f(s(e),w(e))=
	\begin{cases}
		0,& w(e)=0,\\
		w(e)+g(s(e)),& w(e)>0.
	\end{cases}
	\]
	Then $\mathrm{Cost}=\sum_{e\in E} f\big(s(e),w(e)\big)$ satisfies Axioms (1)--(5).
	
	\emph{Axioms (2)--(3).} Layer permutation invariance and edge-wise decomposability hold because $f$ depends only on the counts $(s(e),w(e))$ and the total cost is a sum over edges.
	
	\emph{Axiom (1).} With unit weak cost, each weak instance contributes $1$: when $w(e)>0$ the term $w(e)$ adds exactly one per weak label, and when $w(e)=0$ the cost is $0$.
	
	\emph{Axiom (4).} If $w(e)=0$ there is no strong/weak conflict, so $f(s,0)=0$. If $w(e)>0$, the disagreement surcharge is $g(s(e))$, which is non-decreasing in $s(e)$ and satisfies $g(0)=0$.

	\emph{Axiom (5).} Define the disagreement surcharge at counts $(s(e),w(e))$ by
	$\delta(s(e),w(e)):=f(s(e),w(e))-w(e)$ (the part beyond the unit weak-instance cost).
	Duplicating a layer $l$ where $e\in E_l$ is labeled weak changes $(s(e),w(e))$ to $(s(e),w(e)+1)$, and Axiom (5)
	requires $\delta(s(e),w(e)+1)=\delta(s(e),w(e))$ for all $w(e)\ge 1$.
	For our $f$, $\delta(s(e),0)=0$, while for $w(e)>0$,
	\[
	\delta(s(e),w(e))=f(s(e),w(e))-w(e)=(w(e)+g(s(e)))-w(e)=g(s(e)),
	\]
	which is independent of $w(e)$. Hence duplication leaves the disagreement part unchanged:
	\[
	f(s(e),w(e)+1)-(w(e)+1)=g(s(e))=f(s(e),w(e))-w(e).
	\]
\end{proof}

\begin{proof}[Proof of \Cref{lemma:eqgeneral}]
	Let $S_1^*, \ldots, S_k^*$ be the initial labelings of strong edges.
	Let $F$ be the multiset of all strong edge instances in $S^*$ that participate in a disagreement. 
	That is, $e \in F_i = F \cap E_i$ if $e \in S_i^*$ (edge $e$ is strong in layer $i$) and there exists some layer $j \neq i$ such that $e \in E_j \setminus S_j^*$ (edge $e$ is weak in layer $j$). 
	The size $|F| = \sum_{i \in [k]} |F_i|$ represents the total count of such strong edge instances involved in disagreements.
	By definition, the disagreement score $d_k(S_1^*,\ldots,S_k^*) = \sum_{e \in E} \id{e} \cdot s(e)$, where $s(e)$ is the number of layers in which edge $e$ is strong, and $\id{e}=1$ if $e$ is weak in at least one layer. This sum precisely counts each strong instance of an edge that is part of a disagreement. Therefore, $d_k(S_1^*,\ldots,S_k^*) = |F|$.
	
	Let $W_i^* = E_i \setminus S_i^*$ be the set of weak edges in layer $i$ for the initial labeling $S^*$. The objective value for $S^*$ is
	$C^* = \sum_{i\in[k]} |W_i^*| + d_k(S_1^*,\ldots,S_k^*) = \sum_{i\in[k]} |W_i^*| + |F|$.
	
	We construct a new set of strong edge labelings $H_1^*, \ldots, H_k^*$ by setting $H_i^* = S_i^* \setminus F_i$ for all $i \in [k]$.
	Consider an edge $e$. If $e$ was not involved in any disagreement in $S^*$, its labeling remains unchanged in $H^*$. If $e$ was involved in a disagreement in $S^*$, then all its strong instances (i.e., $e \in F_i$ for relevant $i$) are made weak in $H^*$. Thus, in $H^*$, such an edge $e$ is now weak in all layers where it was previously strong and involved in $F$, and it remains weak in layers where it was already weak. 
	Consequently, the labeling $H^*$ is consistent, which means $d_k(H_1^*,\ldots,H_k^*) = 0$.
	
	Let $W'_i = E_i \setminus H_i^*$ be the set of weak edges in layer $i$ for the new labeling $H^*$.
	Then $W'_i = E_i \setminus (S_i^* \setminus F_i)$. Since $F_i \subseteq S_i^*$, this can be written as $W'_i = (E_i \setminus S_i^*) \cup F_i = W_i^* \cup F_i$.
	As $F_i$ contains edges that were strong in $S_i^*$ (and thus not in $W_i^*$), the sets $W_i^*$ and $F_i$ are disjoint for each layer $i$.
	Therefore, the total number of weak edges in the labeling $H^*$ is
	$\sum_{i\in[k]} |W'_i| = \sum_{i\in[k]} |W_i^* \cup F_i| = \sum_{i\in[k]} (|W_i^*| + |F_i|) = \left(\sum_{i\in[k]} |W_i^*|\right) + |F|$.
	Thus, the objective value for $H^*$ for \textsc{MinMultiLayerSTC} is
	\begin{align*}
		\sum_{i\in[k]} |W'_i| + d_k(H_1^*,\ldots,H_k^*) &= \bigg(\sum_{i\in[k]} |W_i^*|\bigg) + |F|  + 0=C^*. 
	\end{align*}
\end{proof}

\begin{proof}[Proof of \Cref{theorem:approxstc}]
	By \Cref{lemma:eqgeneral}, it suffices to consider consistent labelings when solving \textsc{MinMultiLayerSTC}.
	Let $G=(V, E_1, \ldots, E_k)$ be the multilayer graph and $W=(V_W, E_W)$ the combined wedge graph constructed in \Cref{alg:klayerstc}, where each node $n_e \in V_W$ corresponds to a unique edge $e \in \bigcup_{j} E_j$, and $(n_{e_1}, n_{e_2}) \in E_W$ if $e_1$ and $e_2$ form an open wedge in some layer. Node weights $\omega(n_e)$ are set to the number of layers in which $e$ appears. 
	We reduce \textsc{MinMultiLayerSTC} to a weighted vertex cover (MWVC) problem on $W$:  
	\begin{itemize}[leftmargin=1.5em]
		\item Given a vertex cover $C \subseteq V_W$, label each $e$ as \emph{weak} in all layers if $n_e \in C$, and \emph{strong} otherwise. Since $C$ covers all edges in $W$, every open wedge is blocked by a weak edge, ensuring the STC property in all layers. The cost of the labeling is $\sum_{n_e \in C} \omega(n_e)$.
		\item Conversely, any consistent STC solution $(S_1^*, \ldots, S_k^*)$ defines a vertex cover $C^* = \{n_e \mid e \text{ is \emph{weak} in all layers}\}$. If $C^*$ were not a cover, some wedge would remain with both edges labeled \emph{strong}, violating STC.
	\end{itemize}
	
	Thus, solving MWVC on $W$ is equivalent to solving \textsc{MinMultiLayerSTC} under the consistency constraint. An $\alpha$-approximation for MWVC yields an $\alpha$-approximation for \textsc{MinMultiLayerSTC}.
	
	Constructing $W$ takes $\mathcal{O}(k n m)$ time, as each layer contributes $\mathcal{O}(n m)$ wedges.
	Assigning weights and extracting the final labeling is asymptotically dominated. 
	Thus, the total running time is in $\mathcal{O}(k nm + T_{\text{MWVC}})$.
\end{proof}

\begin{proof}[Proof of \Cref{lemma:eqgeneralplus}]
	The proof follows the same line of reasoning as the proof of \Cref{lemma:eqgeneral}.	
	If an optimal solution $S^*$ (with new edges $E^{N}$) has disagreements $d_k(S_1^*, \ldots, S_k^*) > 0$, we can construct a consistent solution $H^*$ by changing all \emph{strong} instances of disagreeing edges in $S^*$ to \emph{weak}. This increases the count of \emph{weak} edges by exactly $d_k(S_1^*, \ldots, S_k^*)$ while reducing the disagreement penalty to zero, resulting in the same objective value for $H^*$ while the STC properties are preserved.
\end{proof}

\begin{proof}[Proof of \Cref{lemma:doubleapprox}]
	Due to \Cref{lemma:eqgeneralplus}, we know that there is a (optimal) solution for the problem that does not contain any disagreement.
	Assume that there exists an $\alpha$-approximation for \textsc{ConstraintMmlSTC+}.
	Namely, we can find a solution for \textsc{ConstraintMmlSTC+} that is at most $\alpha$ times worse than the optimal solution. 
	Now, for solving \Cref{def:consistentmultistcplus}, if we insert a new \emph{weak} edge in one layer for its corresponding wedge then we have to insert the same \emph{weak} edge in all the layers $\mathcal{L}\subseteq [k]$ in which a corresponding wedge exists (due to (3) in \Cref{def:consistentmultistcplus}). 
	However, it might not be required that the corresponding wedge gets closed by the new \emph{weak} edge in each such layer. 
	To be specific, adding a new \emph{weak} edge in all layers $\mathcal{L}$, 
	might not improve the labeling for those layers. 
	Therefore, we might already have a \emph{weak} edge for the corresponding wedge, thus by adding an additional \emph{weak} edge, we have two \emph{weak} edges, i.e., in total the number of \emph{weak} edges is at most doubled. Therefore, the solution we get for \textsc{MinMultiLayerSTC+} is at most 2$\alpha$ times worse than the optimal solution.
\end{proof}

\begin{proof}[Proof of \Cref{theorem:approxstcplus}]
	Let $G=(V, E_1, \ldots, E_k)$ be the input multilayer graph. \Cref{alg:klayerstcplus} constructs a combined 3-uniform wedge hypergraph $W=(V_W, E_W)$. Nodes $n_e \in V_W$ represent all possible edges $e \in {V \choose 2}$. Hyperedges $\{n_{vu}, n_{vx}, n_{ux}\} \in E_W$ correspond to open wedges $(v, \{u,x\})$ in any layer $G_i$. The weights $\omega(n_e)$ are defined such that an optimal MWVC on $W$ corresponds to an optimal solution to \textsc{ConstraintMmlSTC+} (\Cref{def:consistentmultistcplus}), assuming consistent labelings as per \Cref{lemma:eqgeneralplus}.
	Specifically, $\omega(n_e)$ for an existing edge $e \in E = \bigcup E_j$ is its multiplicity. For a potential new edge $e' \notin E$, $\omega(n_{e'})$ is the number of layers where adding $e'$ would close an open wedge, implicitly reflecting the cost under constraint (3) of the variant (\Cref{def:consistentmultistcplus}).
	
	The algorithm uses an $\alpha$-approximation solver to find a MWVC $C \subseteq V_W$ on the hypergraph $W$. The solution derived from $C$ thus constitutes an $\alpha$-approximate solution for \textsc{ConstraintMmlSTC+}.
	By \Cref{lemma:doubleapprox}, an $\alpha$-approximation for \textsc{ConstraintMmlSTC+} provides a $2\alpha$-approximation for the original \textsc{MinMultiLayerSTC+} problem (\Cref{def:multistcplus}). Therefore, the solution obtained from $C$ is a $2\alpha$-approximation for \textsc{MinMultiLayerSTC+}.
	
	For the running time analysis, the number of vertices in $W$ is $\mathcal{O}(n^2)$, and the number of hyperedges is $\mathcal{O}(k \cdot nm)$ (as each layer contributes $\mathcal{O}(nm)$ potential wedges). Constructing $W$ takes $\mathcal{O}(k \cdot nm)$. If the MWVC algorithm (e.g., pricing for hypergraphs) runs in time linear to the hypergraph size, $T_{MWVC\_H} = \mathcal{O}(k \cdot nm)$. 
	Thus, the total running time is dominated by $\mathcal{O}(k \cdot nm)$.
\end{proof}

\section{Exact ILP Solutions}\label{section:ilp}

In order to solve our multilayer STC and STC+ variants exactly, we develop corresponding ILP formulations.
First, we define the minimization ILP for \textsc{MinMultiLayerSTC}, where we use binary variables $y_{ij}$ which encode \revision{whether} edge $\{i,j\}\in E$ is \emph{strong} $(y_{ij}=0)$ or \emph{weak} $(y_{ij}=1)$, and $m_{ij}$ \revision{which} denotes the multiplicity of edge $\{i,j\}$, i.e., the number of layers $E_\ell$ such that $\{i,j\} \in E_\ell$. 
We can define the ILP as follows because \Cref{lemma:eqgeneral} ensures the existence of an optimal solution without disagreements:
\begin{align}
	&\min \sum_{\{i,j\}\in E}y_{ij} \cdot m_{ij} \\
	&\text{s.t.} \quad y_{ij} + y_{ih} \geq 1 \quad \forall~(i,\{j,h\})\in \mathcal W(E^\ell) \text{ and } \ell \in [k]\\
	&\quad\quad y_{ij}\in\{0,1\}\quad \forall~\{i,j\} \in E.
\end{align}

For the case of the \textsc{MinMultiLayerSTC+} problem, 
we first introduce a binary quadratic program which then can be linearized. 
We use binary variables $y^\ell_{ij}$ which encode \revision{whether} a (possibly new) edge $\{i,j\}\in E_\ell\cup E_\ell^+$ is \emph{strong} $(y_{ij}^\ell=0)$ or \emph{weak} $(y_{ij}^\ell=1)$.
Additionally, we use variables $u_{ij}^\ell$ that encode \revision{whether} edge $\{i,j\}$ exists in layer $\ell\in[k]$ with $u_{ij}^\ell=1$ if edge $\{i,j\}\in E_\ell\cup E_\ell^+$, and $u_{ij}^\ell=0$, otherwise. 
Clearly, the product $u_{ij}^\ell y_{ij}^\ell$ is one if and only if edge $\{i,j\}\in E_\ell\cup E_\ell^+$ will be a \emph{weak} edge in the final result, leading to 
the quadratic program:
\begin{align}
	&\min \sum_{\ell\in [k]}\sum_{ij\in {V \choose 2}} u_{ij}^\ell\cdot y_{ij}^\ell\\
	&\text{s.t.} \quad u_{ij}^\ell y_{ij}^\ell + u_{ih}^\ell y_{ih}^\ell + u_{jh}^\ell y_{jh}^\ell \geq 1 \quad \label{eq:ilp1}\\&\hspace{30mm}\forall~(i,\{j,h\})\in \mathcal W(E_\ell) \text{ and } \ell\in[k]\notag\\
	&\quad\quad y_{ij}^\ell - y_{ij}^h= 0 \quad \forall~\{i,j\} \in \textstyle{V \choose 2} \text{ and } 1\leq \ell \leq h \leq k\label{eq:ilp2}\\
	&\quad\quad u_{ij}^\ell = 1 \quad \forall~\{i,j\}\in E_\ell \text{ and } \ell\in[k]\label{eq:ilp3}\\
	&\quad\quad y_{ij}^\ell \geq u_{ij}^\ell \quad \forall~\{i,j\}\in \textstyle{V \choose 2}\setminus E_\ell \text{ and } \ell\in[k]\label{eq:ilp4}\\
	&\quad\quad y_{ij}^\ell\in\{0,1\}\quad \forall~\{i,j\} \in \textstyle{V \choose 2} \text{ and } \ell\in[k]\\
	&\quad\quad u_{ij}^\ell\in\{0,1\}\quad \forall~\{i,j\} \in \textstyle{V \choose 2} \text{ and } \ell\in[k].
\end{align}
\Cref{eq:ilp1} ensures that for each wedge in any of the layers either one of the wedge edges are \emph{weak}, or a new edge is inserted. Moreover, \Cref{eq:ilp2} guarantees that if an edge is \emph{weak} in one layer, it is \emph{weak} in all layers. \Cref{eq:ilp3} forces all edges that exist in the multilayer graph to exist in the solution. Finally, \Cref{eq:ilp4} ensures that all newly inserted edges are labeled \emph{weak}.
By linearization of the binary products, we obtain the ILP:
\begin{align}
	&\min \sum_{\ell\in [k]}\sum_{ij\in {V \choose 2}} z_{ij}^\ell\\
	&\text{s.t.} \quad z_{ij}^\ell + z_{ih}^\ell + z_{jh}^\ell \geq 1 \quad \forall~(i,\{j,h\})\in \mathcal W(E_\ell) \text{ and } \ell\in[k]\notag\\
	&\quad\quad z_{ij}^\ell \leq y_{ij}^\ell \quad \{i,j\} \in \textstyle{V \choose 2} \text{ and } \ell\in[k]\\
	&\quad\quad z_{ij}^\ell \leq u_{ij}^\ell \quad \{i,j\} \in \textstyle{V \choose 2} \text{ and } \ell\in[k]\\
	&\quad\quad z_{ij}^\ell \geq y_{ij}^\ell + u_{ij}^\ell -1 \quad \{i,j\} \in \textstyle{V \choose 2} \text{ and } \ell\in[k]\\
	&\quad\quad y_{ij}^\ell - y_{ij}^h= 0 \quad \forall~\{i,j\} \in \textstyle{V \choose 2} \text{ and } 1\leq \ell \leq h \leq k\\
	&\quad\quad u_{ij}^\ell = 1 \quad \forall~\{i,j\}\in E_\ell \text{ and } \ell\in[k]\\
	&\quad\quad y_{ij}^\ell \geq u_{ij}^\ell \quad \text{for all } \{i,j\}\in \textstyle{V \choose 2}\setminus E_\ell \text{ and } \ell\in[k]\\
	&\quad\quad y_{ij}^\ell\in\{0,1\}\quad \forall~\{i,j\} \in \textstyle{V \choose 2} \text{ and } \ell\in[k]\\
	&\quad\quad u_{ij}^\ell\in\{0,1\}\quad \forall~\{i,j\} \in \textstyle{V \choose 2} \text{ and } \ell\in[k]\\
	&\quad\quad z_{ij}^\ell\in\{0,1\}\quad \forall~\{i,j\} \in \textstyle{V \choose 2} \text{ and } \ell\in[k].
\end{align}

\section{Details of the Datasets}\label{appendix:datasets}
We provide the details of the eight dataset.
\begin{itemize}
	\item \emph{AUCS} ($5$ layers) contains five different online and offline relations (working relationship, leisure activities, regular lunch, co-authorship, and Facebook friendship) among faculty members of the CS department at Aarhus University~\citep{magnani2013combinatorial}.
	
	\item \emph{Hospital} ($5$ layers) is a face-to-face contact network between hospital patients and health care workers~\citep{vanhems2013estimating}, where each layer represents one day of recorded interactions.
	
	\item \emph{Airports} ($37$ layers) is an aviation transport network of European airports~\citep{cardillo2013emergence}, where layers correspond to different airline companies.
	
	\item \emph{Rattus} ($6$ layers) contains different types of genetic interactions of \emph{Rattus norvegicus}~\citep{de2015structural}, with each layer representing a distinct interaction type.
	
	\item \emph{FfTwYt} ($3$ layers) is a multilayer social network linking user accounts across \emph{FriendFeed}, \emph{Twitter}, and \emph{YouTube}~\citep{DickisonMagnaniRossi2016}, where each platform forms one layer.
	
	\item \emph{Knowledge} ($30$ layers) is derived from the \emph{FB15K-237} knowledge graph~\citep{toutanova2015representing}, where nodes are entities and each layer corresponds to a distinct relation type.
	
	\item \emph{HomoSap} ($7$ layers) represents different types of genetic and protein interactions between human genes curated from the BioGRID database~\citep{de2015muxviz}. 
	The layers correspond to: Physical Association, Association, Direct Interaction, Colocalization, Additive Genetic Interaction, Suppression Genetic Interaction, and Synthetic Genetic Interaction. 
	Each layer captures a distinct biological interaction mechanism between genes or proteins.

	\item \emph{DBLP} ($168$ layers) is a subgraph of the DBLP bibliography~\citep{ley2002dblp} restricted to $A$ and $A^*$ ranked conferences (CORE ranking). Each layer corresponds to one conference, and edges represent co-authorships within that venue.
\end{itemize}

\medskip
\revision{Across all datasets, tie strength is defined structurally via STC.
We do not impose domain-specific semantics on strong versus weak ties and we assume that tie strength is semantically comparable across layers.}

\end{document}